\documentclass[twocolumn, 10pt, amsfonts, amsmath, amssymb, superscriptaddress, nofootinbib, prx]{revtex4-2}
\usepackage{mathtools}
\usepackage[dvipsnames]{xcolor}
\usepackage[linkcolor=Blue,citecolor=Blue,urlcolor=Blue,colorlinks=true]{hyperref}
\usepackage{xpatch}
\usepackage{siunitx}

\usepackage[inline]{enumitem}

\usepackage[normalem]{ulem}
\usepackage{xcolor}

\makeatletter
\xpatchcmd{\@ssect@ltx}{\@xsect}{\protected@edef\@currentlabelname{#8}\@xsect}{}{}
\xpatchcmd{\@sect@ltx}{\@xsect}{\protected@edef\@currentlabelname{#8}\@xsect}{}{}
\makeatother

\newcommand*{\fullref}[1]{\hyperref[{#1}]{\ref*{#1} ``\nameref*{#1}''}}

\DeclareMathOperator{\tr}{tr}

\renewcommand{\vec}{\boldsymbol}
\def\mat{\sigma_\text{L}^{}}
\def\eul{\sigma_\text{E}^{}}

\usepackage{graphicx}
\usepackage{dcolumn}
\usepackage{bm}

\begin{document}

\title{Geometry-induced patterns through mechanochemical coupling}

\author{Laeschkir Würthner}%
\thanks{These authors contributed equally to this work.}
\affiliation{%
Arnold Sommerfeld Center for Theoretical Physics (ASC) and Center for NanoScience (CeNS), Department of Physics,\\
Ludwig-Maximilians-Universität München, Theresienstraße 37, D–80333 Munich, Germany
}%
\author{Andriy Goychuk}%
\thanks{These authors contributed equally to this work.}
\affiliation{%
Arnold Sommerfeld Center for Theoretical Physics (ASC) and Center for NanoScience (CeNS), Department of Physics,\\
Ludwig-Maximilians-Universität München, Theresienstraße 37, D–80333 Munich, Germany
}
\affiliation{Present address: Institute for Medical Engineering and Science, Massachusetts Institute of Technology, Cambridge, MA 02139, United States}
\author{Erwin Frey}%
\email{frey@lmu.de}
\affiliation{%
Arnold Sommerfeld Center for Theoretical Physics (ASC) and Center for NanoScience (CeNS), Department of Physics,\\
Ludwig-Maximilians-Universität München, Theresienstraße 37, D–80333 Munich, Germany
}%
\affiliation{%
Max Planck School Matter to Life, Hofgartenstraße 8, D-80539 Munich, Germany
}%

\date{\today}

\begin{abstract}
Intracellular protein patterns regulate a variety of vital cellular processes such as cell division and motility, which often involve dynamic changes of cell shape. 
These changes in cell shape may in turn affect the dynamics of pattern-forming proteins, hence leading to an intricate feedback loop between cell shape and chemical dynamics.
While several computational studies have examined the resulting rich dynamics, the underlying mechanisms are not yet fully understood.
To elucidate some of these mechanisms, we explore a conceptual model for cell polarity on a dynamic one-dimensional manifold. 
Using concepts from differential geometry, we derive the equations governing mass-conserving reaction--diffusion systems on time-evolving manifolds. 
Analyzing these equations mathematically, we show that dynamic shape changes of the membrane can induce pattern-forming instabilities in parts of the membrane, which we refer to as regional instabilities.
Deformations of the local membrane geometry can also (regionally) suppress pattern formation and spatially shift already existing patterns.
We explain our findings by applying and generalizing the local equilibria theory of mass-conserving reaction--diffusion systems.
This allows us to determine a simple onset criterion for geometry-induced pattern-forming instabilities, which is linked to the phase-space structure of the reaction--diffusion system.
The feedback loop between membrane shape deformations and reaction-diffusion dynamics then leads to a surprisingly rich phenomenology of patterns, including oscillations, traveling waves, and standing waves that do not occur in systems with a fixed membrane shape.
Our work reveals that the local conformation of the membrane geometry acts as an important dynamical control parameter for pattern formation in mass-conserving reaction--diffusion systems.
\end{abstract}

\maketitle


\section{Introduction}
\label{sec:introduction}

Many vital processes in living systems, such as cell division, motility, nutrient uptake, and growth, involve dynamic cell shape changes that are driven by forces produced by cytoskeletal structures and membrane-binding proteins. 
Mechanisms for cytoskeleton-induced deformation of the cell membrane are many, and include the polymerization of actin filaments~\cite{Pollard.Borisy2003, Pollard.Cooper2009}, guided by proteins that promote actin nucleation, polymerization and branching~\cite{Welch.Mullins2002}, or the generation of active stresses through myosin motor proteins in the actomyosin cortex~\cite{Shih.Rothfield2006, Chugh.Paluch2018}.
Remarkably, myosin--VI motor proteins can reshape membranes on their own by means of highly curvature-sensitive motor-protein--lipid interactions~\cite{Rogez.etal2019}. 
This molecular feature is akin to proteins containing BAR-domains that can directly induce shape deformations by binding to the cell membrane~\cite{Zimmerberg.Kozlov2006, Prinz.Hinshaw2009, Simunovic.etal2015,McMahon.etal2015, Gov.2018}.
Cells coordinate these different processes by relying on regulatory signalling pathways and spatiotemporal protein organization involving, for example, the eukaryotic Rho family of GTPases which controls actomyosin polymerization and contractility~\cite{Lawson.Ridley2018}.
In addition, spatiotemporal protein patterns arise from an interplay between localized biochemical reactions and diffusive transport~\cite{Turing.1952,Klausmeier.1999, Green.etal2015, Halatek.etal2018}, as well as possibly advective transport~\cite{Goehring.etal2011, Gross.etal2018}.
All these processes show that intracellular reaction--diffusion systems are quite generally able to control cell shape, as was recently demonstrated in a minimal reconstituted setup where the \emph{E. coli} MinDE protein system induced lipid vesicle deformations even in the absence of cytoskeletal proteins~\cite{Litschel.etal2018, Christ.etal2021, Fu.etal2021}.
Since, conversely, cell geometry can guide protein pattern formation~\cite{Thalmeier.etal2016, Gessele.etal2020, Feddersen.etal2021, Wigbers.etal2021, Wuerthner.etal2021,Burkart.etal2022}, this generically gives rise to mechanochemical feedback loops~\cite{Goehring.Grill2013, Hannezo.Heisenberg2019, Goychuk.2019, LeRoux.2021}.

Such mechanochemical coupling implies an intricate interplay between dynamic shape deformations of the membrane, cytoskeletal dynamics, and chemical reaction kinetics.
Theoretical investigations that address this rich topic range from 
computational models~\cite{Miller.etal2018, Tamemoto.Noguchi2020} to models that place greater emphasis on the underlying molecular processes; for reviews see e.g.~Refs.~\cite{Gov.2018, Alimohamadi.Rangamani2018, Frey.Idema2021}.
For example, recent studies have addressed the impact of curved proteins~\cite{Tozzi.etal2019}, phase separation of membrane-binding proteins~\cite{Yuan.etal2021}, actin polymerization~\cite{Fosnaric.etal2019}, and contractility of the actomyosin cortex~\cite{Mietke.etal2019a} on membrane shape dynamics, as well as the interplay between morphogen and tissue dynamics~\cite{Mercker.etal2016}.
While these studies have identified complex behavior such as membrane waves or complex three-dimensional shapes, they mostly rely on numerical simulations. 
Analytical methods like linear stability analysis have been employed to analyze the effect of mechanical and geometrical degrees of freedom on the onset of pattern formation in simple reaction-diffusion systems~\cite{Tamemoto.Noguchi2020,Mietke.etal2019a,Tozzi.etal2019}.
However, a comprehensive theoretical framework for studying the impact of geometric effects on the resulting protein patterns in the fully nonlinear regime is currently lacking.

To further elucidate the theoretical understanding of systems with mechanochemical coupling, we here study a minimal model in the nonlinear regime employing analytical methods.
Specifically, we consider a conceptual model of cell polarity given by a mass-conserving two-component reaction--diffusion system~\cite{Mori.etal2011, Brauns.etal2020} on a one-dimensional manifold, whose shape can evolve dynamically over time.
Compared to a biologically realistic cell polarity model, the simplifications are twofold: 
\begin{enumerate*}[label=(\roman*)]
\item Instead of a complex protein reaction network for cell polarity~\cite{Goryachev.Leda2017, Kluender.etal2013}, we consider a reduced model with one protein species that can diffuse either in the cytosol or on the membrane.
\item The cell membrane is considered as a deformable one-dimensional manifold.
\end{enumerate*}
The membrane shape deformations cause inhomogeneous membrane compressions or dilations, leading to (local) accumulation or dilution of particle densities.
Since protein densities are important control parameters in mass-conserving reaction--diffusion systems~\cite{Halatek.Frey2018, Brauns.etal2020, Frey.Brauns2020, Wuerthner.etal2021}, deformations of the one-dimensional manifold can qualitatively change the dynamics of protein pattern formation.
In turn, if the local density of proteins also drives the dynamics of the one-dimensional manifold, then this leads to a feedback loop between shape changes of the manifold and reaction--diffusion dynamics. 
The goal of our analysis is to uncover important physical mechanisms underlying this intricate coupling between pattern formation and changes in membrane geometry---using a generic model.

We study this generic model by extending the \emph{local equilibria theory}, a recently developed framework for analyzing mass-conserving reaction--diffusion systems~\cite{Halatek.Frey2018, Brauns.etal2020}, to explicitly account for shape deformations of the manifold and the resulting changes in local geometry.
In particular, we focus on two exemplary cases where
\begin{enumerate*}[label=(\roman*)]
    \item the shape is deformed adiabatically by some external agent, and
    \item the local concentration of proteins controls the dynamic shape changes of the membrane, for example by driving local outward growth through actin polymerization~\cite{Gov.Gopinathan2006}. 
\end{enumerate*}
In the latter case, we consider the one-dimensional membrane as a fluid-like boundary under line tension.
Specifically, our model could be interpreted as describing the dynamics of a small cell cortex section whose outward growth is (locally) driven by proteins.
Therefore, to close our set of equations, we assume that proteins generate local forces that drive outward motion of the membrane along the direction of its normal vector.

We begin in Section~\ref{subsec:local_equilibria_theory} by reviewing the \emph{local equilibria theory} in the context of a planar one-dimensional system with fixed geometry.
In Sections~\ref{sec:deforming-geometry}~and~\ref{sec:reference-frame}, we then apply concepts of differential geometry to describe the dynamics of a one-dimensional manifold, which sets the stage for coupling the manifold's geometry to the chemical degrees of freedom.
By invoking the gauge invariance of the number of proteins enclosed in a given control volume with respect to deformations of the membrane geometry, we derive in Section~\ref{sec:time-evol-density-fields} how the number density of proteins responds to the shape dynamics of the one-dimensional manifold.
We then show in Section~\ref{secIII-two-comp-deform-geom} how these general concepts apply to the specific case of a two-component mass-conserving reaction-diffusion system.
In Section~\ref{sec:lateral-instability}, we extend the local equilibria theory to systems that include shape deformation of the manifold and ensuing changes in their geometry.
To test our theoretical results, in Sections~\ref{subsec:external_deform}--\ref{subsec:pattern-interface-shift} we study the response of a two-component mass-conserving reaction--diffusion system to shape deformations that are driven by an external agent.
Finally in Sections~\ref{subsec:dynamic_coupling}~and~\ref{subsec:time_scales}, we couple the conformational dynamics of the membrane to the local density of proteins, and study how such a system self-organizes in space and time.

\section{Local equilibria theory}
\label{subsec:local_equilibria_theory}
\begin{figure*}[t]
\includegraphics{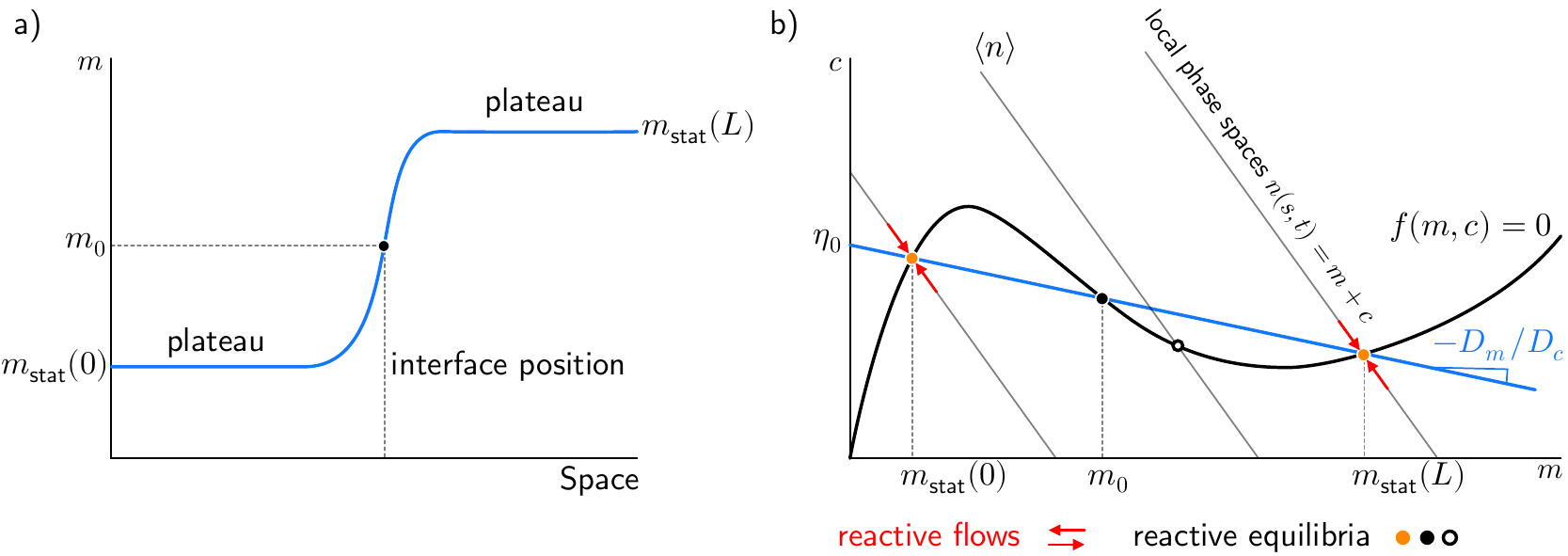}
\caption{%
Stationary pattern of a two-component mass-conserving reaction--diffusion system in phase-space. a) Stationary spatial profile shown for the membrane species $m$, consisting of two plateaus connected by an interface (mesa pattern).
The interface position is defined by the inflection point of the pattern (black filled dot).
b) Phase-space representation of the pattern shown in a).
The stationary solution lies on a linear subspace (flux-balance subspace, blue line), and the slope of this line is given by the ratio of the diffusion coefficients (cf.~Eq.\eqref{eq:twocomp_static_fbs}).
The plateau values are determined by reactive flows (diffusive fluxes are zero in these regions).
In phase-space, the reactive equilibria corresponding to the plateau values (orange filled dots) are given by intersections of local phase spaces (thin gray lines) with the reactive nullcline (black line), and $\langle n \rangle$ is the average total density.
Note that due to mass-conservation, reactive flows (red arrows) are always parallel to the local phase spaces.
}
\label{fig:phase-space}
\end{figure*}

One of the main goals of this work is to find generic principles of pattern formation through reactions and diffusion on manifolds whose conformations change dynamically.
To that end, we build on a recently developed framework for mass-conserving reaction--diffusion (MCRD) systems termed \emph{local equilibria theory}~\cite{Halatek.Frey2018, Brauns.etal2020, Frey.Brauns2020}.
With this framework, one can characterize the dynamics of MCRD systems by analyzing the phase portrait.
However, since the local equilibria theory was originally developed in the context of a fixed spatial domain, it is not clear a priori how to apply it to a system where the patterns emerge on a manifold whose conformation and hence internal geometry changes over time.
Interestingly, through our analysis we find that the main concepts of the local equilibria theory carry over to MCRD systems on dynamic manifolds without major modifications.
Before we proceed with this main subject of our work, we first recapitulate the key points of local equilibria theory.

The basic idea is to think of a spatially extended system as being decomposed into a set of compartments that are coupled by diffusion. 
For an isolated compartment, one can determine the homogeneous steady state (\emph{local reactive equilibrium}) and its stability, which both depend on the total particle densities within that compartment. 
Since diffusion redistributes these total densities, the local reactive equilibria will shift and change over time.
The local reactive equilibria inside each compartment then serve as a scaffold for the spatially extended system, which allows one to study pattern formation by performing a phase portrait analysis.

To illustrate these ideas with a concrete example, consider a two-component MCRD system consisting of one protein species which can cycle between a membrane-bound state $m$ and a cytosolic state $c$. 
On a static one-dimensional domain, e.g., an arbitrary curve in space, the dynamics is given by the following two-component MCRD model:
\begin{subequations}
\label{eq:twocomp_static}
\begin{align}
    \partial_t m(s,t)
    &=D_m \partial_s^2 m +f(m,c) \label{eq:twocomp_static_a}
    \, , \\
    \partial_t c(s,t)&=D_c \partial_s^2 c -f(m,c) 
    \, . 
\label{eq:twocomp_static_b} 
\end{align}
\end{subequations}
Here, $s$ denotes the arc length while the reaction term $f\left(m,c\right)$ describes the local attachment and detachment kinetics of the protein species.
In Appendix~\ref{appendix:reaction_term} we provide the reaction kinetics that was used in this study. 
However, we emphasize that the main conclusions in this work do not depend on the specific choice of $f(m,c)$, as will become clear in the following sections~\cite{Brauns.etal2020, Frey.Brauns2020}.

The dynamics of Eq.~\eqref{eq:twocomp_static} conserves the total average density of proteins:
\begin{equation}
    \langle n \rangle 
    \coloneqq 
    \frac{1}{L}\int_{0}^L ds \, n\left(s,t\right)
    \, ,
\end{equation}
where ${n\left(s,t\right)=m\left(s,t\right)+c\left(s,t\right)}$ describes the local total protein density and $L$ is the length of the line.
Since the reaction kinetics conserves the total density, the reactive flow in phase space must point in the direction of \emph{local reactive phase spaces} given by ${n(s,t)=m(s,t)+c(s,t)}$ (Fig.~\ref{fig:phase-space}). 
The intersections between the local reactive phase spaces and the \emph{reactive nullcline}, obtained from the equation ${f(m,c)=0}$, determine the local reactive equilibria $\left(m^\ast(n),c^\ast(n)\right)$. 
Hence, the values of the local reactive equilibria and how they change depend on the shape of the reactive nullcline and the total density $n$.
This further implies that for a given system (specified by $f(m,c)$), the total density $n$ plays the role of a control parameter for the local dynamics. 

The dynamics of $n(s,t)$ is driven by diffusion, which can be shown by adding Eqs.~\eqref{eq:twocomp_static_a} and~\eqref{eq:twocomp_static_b}:
\begin{align}
\label{eq:twocomp_static_totalmass}
    \partial_t n(s,t) 
    &= 
    D_c \, 
    \partial_s^2
    \left[c(s,t)
    +
    \frac{D_m}{D_c} \, m(s,t)
    \right] \nonumber \\
    &:= - \partial_s j(s,t) 
    \, ,
\end{align}
where the diffusive density flux $j(s,t)$ is given by a combination of cytosolic and membrane density gradients ${j(s,t)=-D_c \partial_s c(s,t) - D_m  \partial_s m(s,t)}$.
From the dynamics of the total density, see Eq.~\eqref{eq:twocomp_static_totalmass}, one directly infers that any stationary pattern, $m_\text{stat}(s)$ and $c_\text{stat}(s)$, must be constrained to a linear subspace in phase space that is (for no-flux or periodic boundary conditions) determined by:
\begin{equation}
\label{eq:twocomp_static_fbs}
    c_\text{stat}(s) 
    + 
    \frac{D_m}{D_c} \, m_\text{stat}(s) 
    = 
    \eta_0
    \, ,
\end{equation}
where $\eta_0$ is a constant of integration. 
The linear subspace, given by Eq.~\eqref{eq:twocomp_static_fbs}, is termed the \emph{flux-balance subspace} (FBS) and states that in steady state the diffusive fluxes in $m$ and $c$ must be balanced such that the net flux is zero (see Eq.~\eqref{eq:twocomp_static_totalmass}).

As shown in Ref.~\cite{Brauns.etal2020}, the condition for the establishment of spatial density patterns is linked to the slope of the reactive nullcline by a simple geometric criterion in phase space:
a homogeneous steady state becomes unstable to spatial perturbations (laterally unstable) when the slope of the reactive nullcline $s_\text{nc}(n)$ is steeper than the slope of the FBS:
\begin{equation}
\label{eq:secII-twocomp_static_slope-criteria}
    s_\text{nc}(n)
    =
    \partial_m\left.c^\ast(m)\right|_n 
    < 
    -\frac{D_m}{D_c}
    \, .
\end{equation}
The underlying mechanism of this instability lies in a coupling between mass-redistribution and reactive flows: regions with a high density of membrane-bound proteins act as a sink for cytosolic particles due to (nonlinear) attachment to the membrane, leading to depletion of cytosolic particles. 
Conversely, regions with a low density of membrane-bound proteins act as a source of cytosolic particles due to detachment from the membrane and hence increase the cytosolic density. 
Redistribution of mass through diffusion further amplifies this effect and leads to a feedback loop between mass-redistribution and reaction kinetics. 
This instability is hence termed the \emph{mass-redistribution instability} and is generic to mass-conserving reaction-diffusion systems.

In Ref.~\cite{Brauns.etal2020} it was furthermore shown that the condition for a lateral instability, see Eq.~\eqref{eq:secII-twocomp_static_slope-criteria}, can be generalized to partitions of the geometry. 
In short, one can dissect a spatial pattern into spatially distinct regions, and associate each region with a regional phase space. 
Viewing these regions in isolation from the rest, one can repeat the same analysis for each region separately and thus reconstruct the global pattern by determining the \emph{regional instability} from Eq.~\eqref{eq:secII-twocomp_static_slope-criteria}.
For a comprehensive discussion of the local equilibria theory and the two-component MCRD model, we refer to Ref.~\cite{Brauns.etal2020}.

\section{reaction--diffusion dynamics on a deforming manifold}
\label{secIII:deforming-interface}
Now that we have recapitulated local equilibria theory for pattern-forming systems on a given spatial domain, we will extend it towards systems on dynamic manifolds.
To that end, we proceed with the following steps.
We start by  providing a generic description of a manifold in terms of curvilinear coordinates, where we restrict the discussion to a one-dimensional system (line).
For a general review of differential geometry of surfaces, we refer to Ref.~\cite{Deserno.2015}.
To determine the time evolution of patterns, we require that their dynamics is independent of the reference frame and that the dynamics of the manifold conserves the number of particles.
By doing so, we derive a set of governing partial differential equations that describes mass-conserving reaction--diffusion systems on a deforming manifold in the laboratory frame.
An important aspect of the dynamics is, that virtually every deformation of the manifold will inevitably change the local density of particles on the manifold.
\begin{figure}[t]
\includegraphics{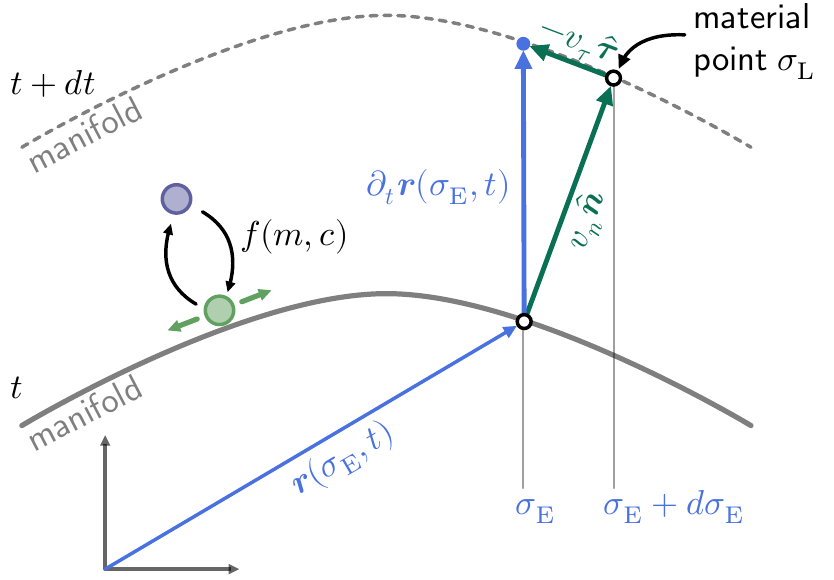}
\caption{%
Conceptual description of a time-evolving manifold.
The line (solid thick gray) can be parameterized either with respect to a stationary ambient coordinate system (\textit{Eulerian coordinates} $\eul$, blue colors) or by using material coordinates $\mat$ (\textit{Lagrangian coordinates} that label points traveling along trajectories normal to the line (black hollow dots). 
As illustrated, the coordinates $\eul$ change over time in the material frame, resulting in a flow ${\eul \to \eul + d \eul}$ of these coordinates that is directed along the tangential part of the velocity vector $\partial_t \vec{r}(\eul,t)$.
We investigate a conceptual two-component mass-conserving reaction--diffusion system on such dynamic manifolds (schematically illustrated by the symbols in purple and green, which represent a cytosolic species $c$ and a membrane species $m$, both diffusing along the line).
}
\label{fig:parametrization}
\end{figure}

\subsection{Describing a deforming geometry}
\label{sec:deforming-geometry}

\begin{figure}[t]
\includegraphics{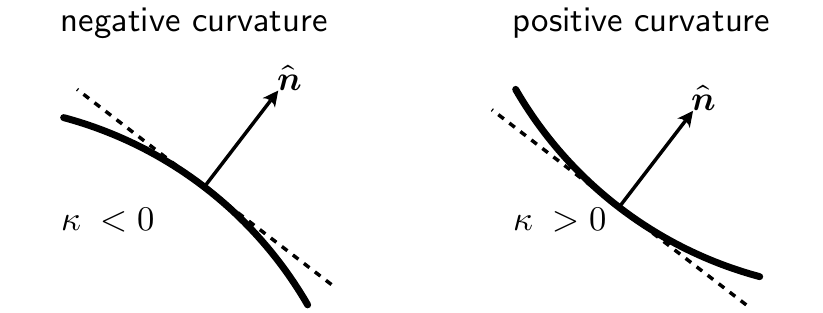}
\caption{%
Convention for the sign of the curvature.
If the manifold (thick black line) curves away from its unit normal vector (black arrow), then the curvature is negative (left panel).
If the geometry (thick black line) curves towards its unit normal vector (black arrow), then the curvature is positive (right panel).
The dashed line represents a flat geometry with vanishing curvature.
}
\label{fig:curvature-sign-convention}
\end{figure}

We begin with the general description of a one-dimensional time-dependent manifold (line). 
We parameterize the line by a time-dependent position vector ${\vec{r}(\sigma, t) \in \mathbb{R}^2}$, where $\sigma$ is an arbitrary curve parameter that labels positions along the line (see Fig.~\ref{fig:parametrization}).
Given a specific parameterization of the position vector, one can then define further geometric features of the line.
The tangent vector of the curve is given by
\begin{equation}
    \vec{\tau}(\sigma, t) = \partial_{\sigma}\vec{r}(\sigma, t) 
    \, .
    \label{eq:secIII-tangent_vector}
\end{equation}
For calculations, it is convenient to consider the normalized tangent vector, which is given by
\begin{equation}
    \hat{\vec{\tau}}(\sigma, t) 
    = 
    \frac{\vec{\tau}(\sigma, t)}{\sqrt{g(\sigma, t)}} 
    \, ,
    \quad\text{where}\quad
    g(\sigma, t) 
    = 
    \lVert\vec{\tau}(\sigma, t) \rVert^2
    \label{eq:secIII-metric}
\end{equation}
refers to the \textit{metric} or \emph{first fundamental form}, which allows one to define arc distances along the curve:
\begin{equation}
    s\left(\sigma,t\right)
    =
    \int_{0}^{\sigma}d\sigma' \, \sqrt{g(\sigma',t)}
    \, .
    \label{eq:secIII-arclength}
\end{equation}
The conformation of the curve is described by the curvature $\kappa$ and formally given by the definition
\begin{equation}
    \partial_s \hat{\vec{\tau}} = \kappa \, \hat{\vec{n}} \,,
    \label{eq:curvature}
\end{equation}
where $\hat{\vec{n}}$ is the unit normal vector on the curve.
Note that the direction of the unit normal vector $\hat{\vec{n}}$ and the sign of the curvature are not uniquely defined but a matter of convention.
Here, we choose the convention that the curvature is negative for a sphere (circle in two-dimensions).
This means that the line \emph{curves away from} its unit normal vector in the case of negative curvature, and towards its unit normal vector in the case of positive curvature (see Fig.~\ref{fig:curvature-sign-convention}).

The dynamics of the curve is determined by its velocity vector $\partial_t \vec{r}$ and can, in the most general case, be decomposed into parts normal and tangential to the curve:
\begin{equation}
    \partial_t\vec{r}(\sigma,t)=v_n\vec{\hat{n}}+v_\tau\vec{\hat{\tau}},
    \label{eq:secIII-velocity-vector-general}
\end{equation}
where $v_n$ and $v_\tau$ refer to the normal velocity and tangential velocity, respectively.
Note that only the normal velocity $v_n$ affects the shape of the curve, while the shape is invariant to deformations in the tangential direction.
In addition, the normal component of the velocity vector can generally be an arbitrarily complex function of the position vector, curvature, and other field variables. 
Hence, the exact form of the normal velocity must be determined by the specific physical system being studied.
Conversely, as we will show in the next section, the tangential velocity strongly depends on the concrete choice of the curve parameterization and on the frame of reference.

\subsection{Frame of reference}
\label{sec:reference-frame}

Up to this point, we have not specified the exact frame of reference for the parameterization of the geometry.
In the following, by analogy with continuum mechanics, we distinguish between two different frames of reference, the \emph{Lagrangian frame} and the \emph{Eulerian frame}.
In the Lagrangian frame, we exploit the fact that tangential motion, i.e., sliding of the curve along its own contour, does not change the shape of a line.
Therefore, we specify the curve parameter $\mat$ to label material points with vanishing tangential velocity along the line.
These material points then move with the line along the normal direction $\vec{\hat{n}}$ (Fig.~\ref{fig:parametrization}).
To further emphasize the special properties of the Lagrangian or co-moving frame, we define the \emph{material derivative} $\mathcal{D}_t$.
If any quantity, such as the position vector or some density field on the one-dimensional manifold, is parameterized by material coordinates $\mat$, then the material derivative is identical to the (local) partial time derivative~\cite{Grinfeld.2009,Grinfeld.2012}
\begin{equation}
    \mathcal{D}_t \coloneqq \left.\partial_t\right|_{\vec{r}(\mat,t)}.
    \label{secIII-material-derivative-definition}
\end{equation} 
This further implies that the material coordinates $\mat$ are time-invariant in the Lagrangian frame of reference, as they should be.

In the \textit{Eulerian frame}, one defines a curve parameter $\eul$ that labels points on the curve which do not move with the material coordinates, i.e., do not flow solely along the normal but also along the tangential direction of the moving line.
The parameterization in this case is given with respect to an ambient coordinate system (\emph{laboratory} frame) (Fig.~\ref{fig:parametrization}) and therefore specifies a fixed coordinate for the position of the line over time (for fixed $\eul$).
Note, however, that while the material coordinates $\mat$ do not change in the Lagrangian frame, they do become time-dependent in the Eulerian frame $\mat=\mat(\eul,t)$ (see Fig.~\ref{fig:parametrization}). 
This is analogous to fluid mechanics, where the coordinates of a fluid parcel in the material frame are fixed over time, while the fluid parcel moves at a certain velocity as seen by an observer in the laboratory frame.

Now, for a consistent physical description that is independent of the exact definition of the laboratory frame (i.e., parameterization in the Eulerian frame), the total time derivative of an arbitrary physical quantity in the laboratory frame must match the time derivative in the Lagrangian frame (i.e., using the operator $\mathcal{D}_t$).
To understand this, we apply the material derivative $\mathcal{D}_t$ to the position vector field $\vec{r}(\eul(\mat,t),t)$ parameterized by Eulerian coordinates $\eul$:
\begin{align}
    \mathcal{D}_t \vec{r}(\mat, t) 
    &= \frac{d}{dt} \vec{r}(\eul(\mat,t), t) \,, \nonumber \\ 
    &= \partial_t \vec{r}(\eul, t) + \partial_{\eul} \vec{r}(\eul, t) \, \partial_t \eul(\mat,t) \, ,\nonumber \\ \label{eq:secIII-material-deriv-position-vector}
    &= \partial_t \vec{r}(\eul, t) + \hat{\vec{\tau}} \,\sqrt{g} \, \partial_t\eul(\mat,t) \, \, , 
\end{align}
where $\partial_t \eul(\mat,t)$ denotes an ambient \emph{coordinate flow} as seen by an observer in the material frame at $\mat$ (see Fig.~\ref{fig:parametrization}).
Note that, because we have chosen the material coordinates $\mat$ such that they label points with vanishing tangential velocity along the line, one finds that ${\mathcal{D}_t \vec{r}(\mat,t)=v_n \vec{\hat{n}}}$; i.e., the material derivative only contains a part normal to the line.
The ambient coordinate flow can be determined by a simple geometric construction (see Fig.~\ref{fig:parametrization}):
\begin{equation}
    v_n dt \,\vec{\hat{n}}-\partial_t \vec{r}(\eul,t)\, dt=d\eul \vec{\tau}
    \, .
    \label{eq:secIII-coordflow-geometric-construction}
\end{equation}
Inserting Eq.~\eqref{eq:secIII-velocity-vector-general} into Eq.~\eqref{eq:secIII-coordflow-geometric-construction} one obtains ${d\eul/dt=-v_\tau/\sqrt{g}}$.
With this result, one can rewrite Eq.~\eqref{eq:secIII-material-deriv-position-vector} to obtain the equivalent form:
\begin{equation}
    \mathcal{D}_t \vec{r}(\mat, t) =v_n\vec{\hat{n}}= \partial_t \vec{r}(\eul, t) - v_\tau\, \hat{\vec{\tau}} \, .
    \label{eq:secIII-material-deriv-position-vector-final}
\end{equation}
To conclude, by combining Eqs.~\eqref{eq:secIII-material-deriv-position-vector} and~\eqref{eq:secIII-tangent_vector}, we find that the material derivative operator in the Eulerian frame is given by:
\begin{equation}
    \mathcal{D}_t \equiv\partial_t -v_\tau \, \partial_s 
    \,,
    \label{eq:secIII-material-derivative-formal-definition}
\end{equation}
where we have also used ${\partial_\sigma = \sqrt{g(\sigma,t)} \,  \partial_s}$ (cf. Eq.~\eqref{eq:secIII-arclength}). 
This operator serves as a link between the laboratory and material frames, and can be applied to any time-dependent quantity defined on the one-dimensional manifold.

\subsection{Conformational dynamics of the line affects density fields}
\label{sec:time-evol-density-fields}

If the conformation of the line changes with time, then what is the time evolution of a density field that is defined on the one-dimensional manifold?
We proceed by considering an arbitrary scalar field $\varrho$ representing, for instance, the density fields of cytosolic and membrane-bound proteins.
We first define the cumulative number of particles up to some coordinate $\sigma$ along the line,
\begin{align}
    N_\varrho(\sigma, t) 
    &\coloneqq \int_{0}^{s(\sigma, t)} ds' \, \varrho(s',t) 
    \nonumber \\ 
    &= \int_{0}^\sigma d\sigma' \, \sqrt{g(\sigma', t)} \, \varrho(\sigma',t) \, ,
    \label{eq:secIII-total-particle-number-general}
\end{align}
where for brevity of notation we have written ${\varrho(\sigma,t) = \varrho(s(\sigma,t),t)}$ using the same symbol for the function.
For mass-conserving systems, the total number of particles remains constant over time, irrespective of any deformation of the line's shape that alters its total or local length.
In contrast, the local density of particles can change as a result of shape deformations, since density fields are defined with respect to the local arc length, which is in general also a time-dependent quantity; see Eq.~\eqref{eq:secIII-arclength}.
To derive the time evolution of density fields, it is therefore useful to first look at the time evolution of the particle number.
Consider the number of particles distributed on an infinitesimal line segment between coordinates $\sigma$ and $\sigma + d \sigma$,
\begin{equation}
    d N_\varrho(\sigma, t)  = d \sigma \, \sqrt{g(\sigma,t)} \, \varrho(\sigma, t) \, ,
    \label{eq:secIII-local-particle-number}
\end{equation}
and further assume that we have chosen a parameterization in the Lagrangian frame, i.e., we set ${\sigma \equiv \mat}$.
The time evolution of the number of particles then follows by applying the material derivative to Eq.~\eqref{eq:secIII-local-particle-number}:
\begin{multline}
    \mathcal{D}_t\bigl[d N_\varrho(\mat, t)\bigr] = d \mat \, \sqrt{g} \, \left[\frac{\mathcal{D}_t g}{2g} \, \varrho + \mathcal{D}_t \varrho(\mat, t)\right] \, ,
    \label{eq:secIII-time-deriv-particle-number-lagrange}
\end{multline}
Since the left-hand side of Eq.~\eqref{eq:secIII-time-deriv-particle-number-lagrange} can only change due to local reactions, $f_\varrho(\mat, t)$, and particle fluxes $j(\mat,t)$ across the boundaries of the line segment $d \mat$, it must be given by the transport equation:  
\begin{multline}
    \mathcal{D}_t\bigl[d N_\varrho(\mat, t)\bigr] = j(\mat, t) - j(\mat + d\mat, t)\\+d \mat \, \sqrt{g(\mat, t)} \, f_\varrho(\mat, t) \, .
    \label{eq:secIII-time_deriv_comoving_reactions}
\end{multline}
The temporal evolution of the metric can be determined from the definition~\eqref{eq:secIII-metric}:
\begin{align}
    \mathcal{D}_t g(\mat,t) 
    &= \mathcal{D}_t \left[\partial_{\mat}\,\vec{r}\right]^2
    = 2 [\partial_{\mat}\mathcal{D}_t\vec{r}]\cdot[\partial_{\mat}\vec{r}]\nonumber\\
    &= 2g\left[\partial_s v_n \vec{\hat{n}}+v_n\partial_s\vec{\hat{n}}\right]\cdot\vec{\hat{\tau}}\nonumber\\
    &= -2\,g\,\kappa\,v_n \,,
    \label{eq:secIII-comoving-timederiv-metric}
\end{align}
where we have used the relation ${\partial_{\mat}=\sqrt{g}\,\partial_s}$, see Eq.~\eqref{eq:secIII-arclength}, and the fact that ${\partial_s \vec{\hat{n}}=-\kappa \vec{\hat{\tau}}}$.
Combining Eqs.~\eqref{eq:secIII-time-deriv-particle-number-lagrange}--\eqref{eq:secIII-comoving-timederiv-metric}, we obtain the governing equation for the density field in the Lagrangian frame:
\begin{equation}
    \mathcal{D}_t\varrho(\mat,t)
    =
    -\partial_s j(\mat,t)
    +f_\varrho(\mat, t) 
    +\kappa \, v_n \,\varrho(\mat,t) 
    \,.
    \label{eq:secIII-time-deriv-density-lagrange}
\end{equation}
The particle flux in Eq.~\eqref{eq:secIII-time-deriv-density-lagrange} can in general include diffusive as well as advective fluxes along the one-dimensional manifold, ${j(\mat,t)=-D_\varrho \, \partial_s \varrho+v_\tau\, \varrho}$, where $D_\varrho$ and $v_\tau$ denote the diffusion coefficient and tangential advection velocity, respectively.
Advective flows along the membrane may be caused, for instance, by spatial heterogeneities in actomyosin contractility~\cite{Mietke.etal2019a} (cortical flows) or relaxation of in-plane elastic stresses of the membrane.
\begin{figure}[!t]
\includegraphics{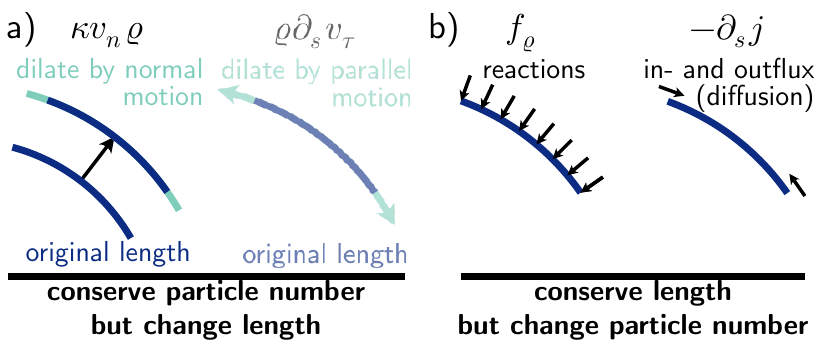}
\caption{%
The different physical contributions to a change in particle concentration on a deforming line. 
a) Mechanisms that conserve the local number of particles on a given line segment while changing the segment's length.
Motion along the normal vector increases (decreases) the length of segments with negative (positive) curvature (cf. Fig.~\ref{fig:curvature-sign-convention}), thus reducing (increasing) the local concentration of particles.
Since we here focus on exterior motion of the line (along a direction normal to the line), we do not consider changes in particle concentrations that would arise due to parallel (interior) motion (faded panel). 
However, one could also account for dilution and accumulation of particle densities due to parallel motion by incorporating a term $\varrho \, v_\tau$ into the density flux in Eq.~\eqref{eq:secIII-time-deriv-density-lagrange}.
b) Reactions and diffusion are mechanisms that change the number of particles on a segment while conserving the segment's length.
}
\label{fig:geometric-contributions}
\end{figure}
In this work, we consider systems where the particles are transported only by diffusion and therefore disregard advective particle fluxes, in line with our choice of reference frame with a vanishing tangential velocity.
The last term in Eq.~\eqref{eq:secIII-time-deriv-density-lagrange} is a purely geometric contribution and accounts for local density variations due to local length extension and contraction.
The various contributions to the local change of the particle density are summarized in Fig.~\ref{fig:geometric-contributions}.

\begin{figure}[t]
\includegraphics{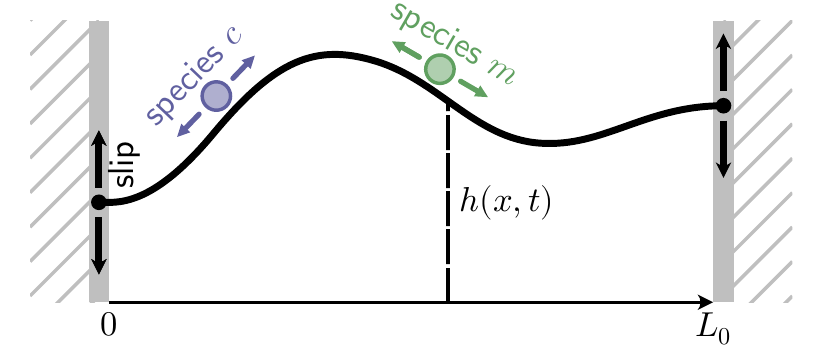}
\caption{%
Illustration of the Monge parameterization. 
The membrane (black solid line) is parameterized by its height relative to a flat line.
The endpoints of the membrane can slip along two solid walls (grey), which are a distance $L_0$ apart.
}
\label{fig:monge-parametrization}
\end{figure}

While the Lagrangian frame is convenient for our analytic calculations, the choice of a specific parameterization in the Eulerian frame allows us to reduce the number of degrees of freedom in our numerical simulations.
To that end, we choose a Monge parameterization of the line contour:
\begin{equation}
    \vec{r}(x,t)=
    \begin{bmatrix}
    x \\
    h(x,t)
    \end{bmatrix} \, ,
\label{eq:secIII-Monge-param}
\end{equation}
where the height field $h(x,t)$ encodes the line conformations, and ${x \in [0,L_0]}$ is the curve parameter (here an Eulerian coordinate ${\eul \equiv x}$).
Thus, by using a Monge parameterization, we eliminate the time evolution of one component of the position vector $\vec{r}(x,t)$, thus retaining only one degree of freedom.
However, since the line is now represented by the graph $h(x,t)$, we explicitly exclude overhangs by using this parametrization.
We further assume that the two opposing endpoints of the membrane are clamped, i.e., forced to a slope of zero, while allowing the line to slip vertically along the boundaries (Fig.~\ref{fig:monge-parametrization}). 
Since the line extends from $x=0$ to $L_0$, we effectively introduce a length constraint, stating that the total length of the membrane may not fall below the minimum distance $L_0$ (Fig.~\ref{fig:monge-parametrization}).

One could generalize this choice to account for overhangs, at the expense of increased model complexity, by introducing additional degrees of freedom that allow the curve to freely move in space (for example described in the Lagrangian frame), and adding physical mechanisms such as stretching rigidity to constrain the total length of the curve.
However, taking mechanical degrees of freedom into account would greatly complicate the dynamics, since stress propagation along lines exhibits a rather intricate dynamics, as was shown for polymers~\cite{Hallatschek.etal2005, Hallatschek.etal2007a,  Hallatschek.etal2007b, Obermayer.Hallatschek2007,Obermayer.etal2007}.
Here, we disregard these additional complexities and focus on the interplay between biochemical pattern formation and the shape  deformation of the line.

Finally, using the definition of the material derivative in Eulerian coordinates, Eq.~\eqref{eq:secIII-material-derivative-formal-definition}, the dynamics of density fields, Eq.~\eqref{eq:secIII-time-deriv-density-lagrange}, can be translated to the laboratory frame:
\begin{multline}
    \partial_t\varrho(\eul,t)=-\partial_s j(\eul,t)+f_\varrho(\eul, t)\\ +\kappa \, v_n \,\varrho(\eul,t)+v_\tau \partial_s\varrho(\eul,t) \,.
    \label{eq:secIII-time_deriv-density-euler}
\end{multline}
This reaction-diffusion equation on a deforming line taken together with the time evolution of the line's shape, Eq.~\eqref{eq:secIII-velocity-vector-general}, fully specify the dynamics of a density field on a manifold that changes its conformation.
Next, our goal is to understand how shape deformations of a line affect protein pattern formation.

\subsection{Two-component MCRD system on dynamically evolving manifolds}
\label{secIII-two-comp-deform-geom}

As we have now established the framework to study reaction--diffusion systems on lines exhibiting conformational dynamics, we proceed with the generic description of a two-component MCRD system on such a line in the laboratory frame.
The governing equations that describe the dynamics of the density fields and the line conformation in Monge parameterization are then derived from Eqs.~\eqref{eq:secIII-velocity-vector-general},~\eqref{eq:secIII-Monge-param}, and~\eqref{eq:secIII-time_deriv-density-euler}:
\begin{subequations}
\label{eq:secIII-two-comp-MCRD-Monge}
\begin{align}
    \partial_t m(x,t) 
    &= 
    D_m \partial_s^2 m + f + \kappa\,v_n\,m + v_\tau\partial_s m 
    \,, 
    \label{eq:secIII-two-comp-MCRD-Monge_a}
    \\
    \partial_t c(x,t) 
    &= 
    D_c \partial_s^2 c - f + \kappa\,v_n\,c + v_\tau\partial_s c
    \,,
    \label{eq:secIII-two-comp-MCRD-Monge_b} 
    \\
    \partial_t h(x,t) 
    &= 
    \sqrt{g(x,t)}\,v_n 
    \,,
\end{align}
\end{subequations}
where ${g(x,t)=1+[\partial_x h(x,t)]^2}$ is the local metric and ${v_\tau=v_n \partial_x h(x,t)}$ denotes the tangential velocity in the Monge parameterization. 
Note that the tangential velocity follows from a geometric construction; see Eq.~\eqref{eq:secIII-coordflow-geometric-construction} and Fig.~\ref{fig:parametrization}.
The second spatial derivative $\partial_s^2$ along the curve corresponds to the (one-dimensional) \emph{Laplace-Beltrami} operator and is explicitly given by:
\begin{align}
\partial_s^2& \equiv\frac{1}{\sqrt{g}}\partial_x\left[\frac{1}{\sqrt{g}}\partial_x\right]=\frac{1}{g}\partial_x^2-\frac{1}{2}\frac{\partial_x g}{g^{2}}\partial_x.
\end{align}
Unlike in the case of a fixed planar geometry, Eqs.~\eqref{eq:twocomp_static_a} and~\eqref{eq:twocomp_static_b}, the two-component mass-conserving reaction--diffusion system on a deforming line, Eqs.~\eqref{eq:secIII-two-comp-MCRD-Monge_a} and~\eqref{eq:secIII-two-comp-MCRD-Monge_b}, is not given in a form where mass-conservation is immediately apparent.
This is due to the fact that the density fields are defined with respect to the local arc length, which is a dynamic quantity itself, as accounted for by the geometric terms $\kappa\,v_n\,m$ and $\kappa\,v_n\,c$ in Eqs.~\eqref{eq:secIII-two-comp-MCRD-Monge_a} and~\eqref{eq:secIII-two-comp-MCRD-Monge_b}.
Therefore, if one allows the conformation of the line to change over time, the total average density is not necessarily conserved by the dynamics.
Instead, the quantity that must be conserved for a system with a mass-conserving reaction dynamics is the total particle number 
\begin{equation}
    N
    =
    \int_0^{L(t)}ds\,(m+c)
    =
    \int_0^{L_0}dx\,
    \sqrt{g(x,t)}\,n(x,t)
    \, .
\end{equation}
This renders our analysis slightly more involved, since the local equilibria theory cannot be applied to the mathematical model, Eqs.~\eqref{eq:secIII-two-comp-MCRD-Monge_a} and~\eqref{eq:secIII-two-comp-MCRD-Monge_b}, in their present form. 

To get around this problem, we consider rescaled densities on the membrane ${\widetilde{m}(x,t):=\sqrt{g(x,t)}\,m(x,t)}$ and in the cytosol ${\tilde{c}(x,t):=\sqrt{g(x,t)}\,c(x,t)}$.
This mapping corresponds to a projection of the line densities along the curve, $m(x, t)$ and $c(x,t)$, onto the parameterization axis (in the case of a Monge representation the $x$-axis), so that
\begin{equation}
\begin{split}
    N 
    &= 
    \int_{0}^{L_0} \!dx \, 
    \sqrt{g(x,t)} \, 
    \big( 
    m(x,t)+c(x, t) 
    \big) 
    \\
    &=
    \int_{0}^{L_0} \!dx \, 
    \big(
    \widetilde{m}(x,t)+\tilde{c}(x, t)
    \big)
    =
    \int_{0}^{L_0} dx \,
    \tilde{n}(x,t) 
    \, .
\label{eq:secIII-density-mapping}
\end{split}
\end{equation}
Thus, $\tilde{n}\,dx$ represents the total number of particles contained within an infinitesimal compartment $dx$.
Using our mapping, one immediately sees that the mapped total average density ${\langle \tilde{n} \rangle = N / L_0}$ is conserved by the dynamics, thus allowing us to apply the local equilibria theory.
The time evolution of the rescaled variables can be determined starting from:
\begin{equation}
    \partial_t \widetilde{m}(x,t)=\sqrt{g}\,\partial_tm(x,t)+\frac{1}{2}\frac{\partial_t g(x,t)}{\sqrt{g(x,t)}}\,m(x,t) \,,
    \label{eq:secIII-mapped-density-derivation}
\end{equation}
and an analogous equation for the cytosolic species $\tilde{c}$.
In the following, only the derivation for the membrane species $\widetilde{m}$ is presented, since the calculations for the cytosolic species $\tilde{c}$ are completely analogous.
The time evolution of the metric $\partial_t g(x,t)$ in the Eulerian frame can be determined similarly as shown in the previous section for $\mathcal{D}_t g(\mat,t)$:
\begin{align}
    \partial_tg(x,t)&=\partial_t\left[(\partial_x \vec{r})\cdot(\partial_x \vec{r})\right]=2\,(\partial_x\partial_t\vec{r})\cdot(\partial_x \vec{r}),\nonumber\\
    &=2\,g\left[\partial_sv_n\vec{\hat{n}}+v_n\partial_s\vec{\hat{n}}+\partial_s v_\tau\vec{\hat{\tau}}+v_\tau\partial_s\vec{\hat{\tau}}\right] \cdot \vec{\hat{\tau}}, \nonumber \\
    &=-2\,g\,\kappa\,v_n+2\,g\,\partial_s v_{\tau},
    \label{eq:secIII-time-deriv-metric-general}
\end{align}
where we used the general expression for the velocity vector Eq.~\eqref{eq:secIII-velocity-vector-general}, the definition Eq.~\eqref{eq:curvature}, and the fact that $\partial_s \vec{\hat{n}}=-\kappa \, \vec{\hat{\tau}}$.
Inserting Eqs.~\eqref{eq:secIII-two-comp-MCRD-Monge_a} and~\eqref{eq:secIII-time-deriv-metric-general} into Eq.~\eqref{eq:secIII-mapped-density-derivation}, we obtain the continuity equation:
\begin{equation}
    \partial_t \widetilde{m}(x,t)=-\partial_x \, j_{\widetilde{m}}^{}(x,t) + \tilde{f}(x,t) \,,
    \label{eq:secIII-mapped-density-membrane}
\end{equation}
where the flux $j_{\widetilde{m}}$ is given by:
\begin{equation}
   j_{\widetilde{m}}(x,t)=-\left[\frac{D_m}{\sqrt{g}}\partial_x\left(\frac{\widetilde{m}(x,t)}{\sqrt{g}}\right) + v_\tau\frac{\widetilde{m}(x,t)}{\sqrt{g}}\right],
   \label{eq:secIII-mapped-flux-membrane}
\end{equation}
and the rescaled reaction term $\tilde{f}(x,t)$ is defined as:
\begin{align}
   \tilde{f}(x,t)&=\sqrt{g(x,t)} \, f(m,c) \nonumber \\
   &=\sqrt{g(x,t)} \, f\left(\frac{\widetilde{m}(x,t)}{\sqrt{g(x,t)}},\frac{\tilde{c}(x,t)}{\sqrt{g(x,t)}}\right) \, .
   \label{eq:secIII-mapped-reaction}
\end{align}
Hence, the mass-conserving dynamics of the rescaled density fields are given by a reaction--diffusion system in conservative form:
\begin{subequations}
\label{eq:secIII-mapped-system}
\begin{align}
    \partial_t \widetilde{m}(x,t)&=-\partial_x \, j_{\widetilde{m}}^{}(x,t)+\tilde{f}(x,t)
    \, ,
    \label{eq:secIII-mapped-system-a} 
    \\
    \partial_t \tilde{c}(x,t)&=-\partial_x \, j_{\tilde{c}}^{}(x,t)-\tilde{f}(x,t)
    \, ,
    \label{eq:secIII-mapped-system-b} 
    \\
    \partial_t h(x,t) &=\sqrt{g(x,t)}\, v_n
    \, .
    \label{eq:secIII-mapped-system-c}
\end{align}
\end{subequations}
Note that the equations governing the dynamics of the rescaled fields are equivalent to a reaction--diffusion system defined on a fixed planar geometry of length $L_0$.
The influence of shape deformations of the line on the reaction--diffusion dynamics is fully absorbed into the metric factor $\sqrt{g(x,t)}$, the conservative fluxes $j_{\widetilde{m}/\tilde{c}}$, and the rescaled reaction term $\tilde{f}(x,t)$.
Strikingly, the reaction term becomes space- and time-dependent, in contrast to the case of a fixed planar geometry~\cite{Brauns.etal2020} where the reaction kinetics has the same form at each point in space and time.
We will show in the following sections that this spatiotemporal inhomogeneity of the reaction term can lead to regional instabilities and thus deformation-induced protein pattern formation.

\subsection{Lateral instability on a deforming manifold}
\label{sec:lateral-instability}
\begin{figure*}[tbh]
\includegraphics[width=0.9\textwidth]{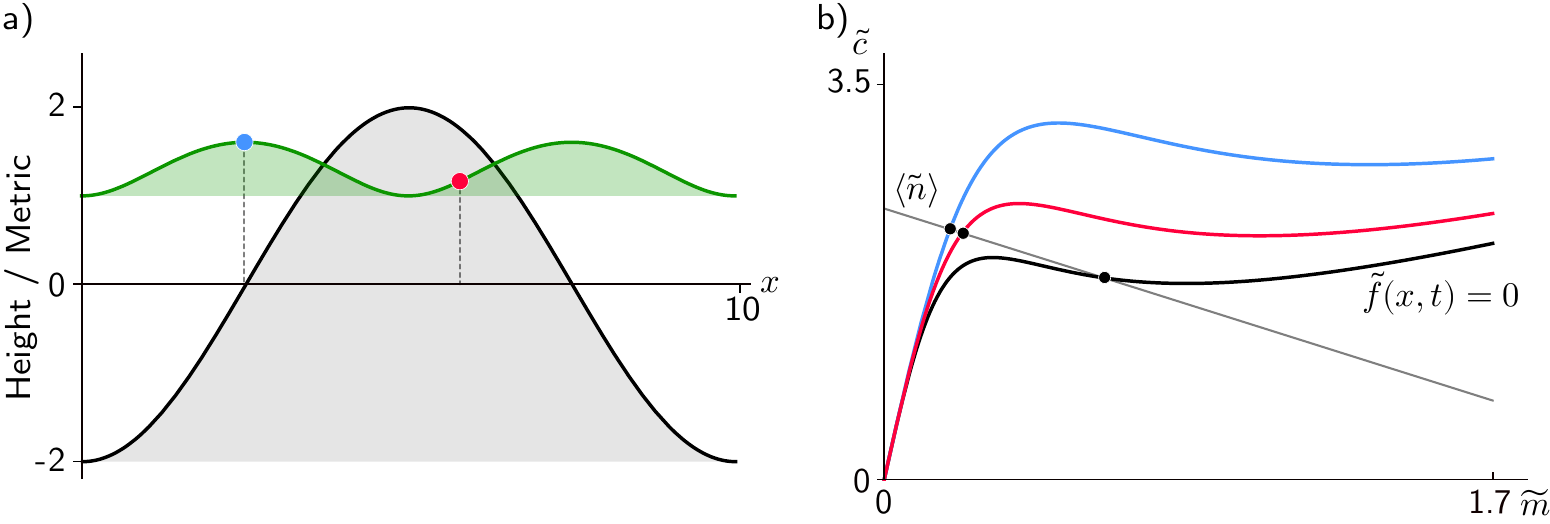}
\caption{%
Shape deformations of the manifold induce spatial inhomogeneities of the reactive nullcline.
a) Height profile (black line) in Monge parametrization and the corresponding spatial profile of the metric (green line) for a given point in time.
b) Phase space geometry for the rescaled variables $\tilde{c}$ and $\widetilde{m}$ at the same time point as in panel a).
The differently colored lines show the reactive nullcline ${f(x,t)=0}$ at different positions in space (as depicted by the blue and red points in panel a)). 
The black line corresponds to the reactive nullcline for a planar geometry, i.e., for regions of the height profile where ${\sqrt{g} \approx 1}$.
The thin gray line shows the local phase space for the (conserved) rescaled total average density $\langle \tilde{n} \rangle$. 
The intersections of this line with the reactive nullclines determines the local equilibria (black dots).
Note that the local phase space intersects the black nullcline in a section where the slope is negative, hence fulfilling the instability criterion~\eqref{eq:secIII-lateral-instability-criterion-mapped}, while the intersections with the blue and red nullclines lie in sections where the slope is positive (laterally stable).
}
\label{fig:nullclines-inhomogeneity}
\end{figure*}

In Sec.~\ref{subsec:local_equilibria_theory} we explained that the homogeneous steady state of the two-component MCRD system (on a fixed planar line) becomes unstable to spatial perturbations when the slope of the reactive nullcline is steeper than the slope of the flux-balance subspace; 
see Eq.~\eqref{eq:secII-twocomp_static_slope-criteria}.
Here we ask whether it is possible to obtain an instability criterion when deformations of the line are considered.

To investigate this question, we start with the description of the two-component MCRD model in the Lagrangian frame, as this is more convenient for analytical calculations, and again project the density fields onto the parameterization axis (note that in this case the curve parameter is $\mat$).
The equations take the same form as for the laboratory frame, see Eq.~\eqref{eq:secIII-mapped-system}, except that there is no tangential drift in the Lagrangian frame (cf. Eq.~\eqref{eq:secIII-mapped-flux-membrane}):
\begin{subequations}
\label{eq:secIII-mapped-system-lagrange}
\begin{align}
\mathcal{D}_t\widetilde{m}(\mat,t)&= \partial_{\mat}\left[\frac{D_m}{\sqrt{g}}\partial_{\mat}\left(\frac{\widetilde{m}}{\sqrt{g}}\right)\right]+\tilde{f}(\mat,t) \,,
\label{eq:secIII-mapped-system-lagrange-a}\\
\mathcal{D}_t\tilde{c}(\mat,t)&= \partial_{\mat}\left[\frac{D_c}{\sqrt{g}}\partial_{\mat}\left(\frac{\tilde{c}}{\sqrt{g}}\right)\right]-\tilde{f}(\mat,t)\, .
\label{eq:secIII-mapped-system-lagrange-b}
\end{align}
\end{subequations}
The dynamics of the (rescaled) local total density $\tilde{n}(\mat,t)$ is obtained by summing Eqs.~\eqref{eq:secIII-mapped-system-lagrange-a} and~\eqref{eq:secIII-mapped-system-lagrange-b}:
\begin{equation}
    \mathcal{D}_t\tilde{n}= \partial_{\mat}\left[\frac{D_m}{\sqrt{g}}\partial_{\mat}\left(\frac{\widetilde{m}}{\sqrt{g}}\right)+\frac{D_c}{\sqrt{g}}\partial_{\mat}\left(\frac{\tilde{c}}{\sqrt{g}}\right)\right] \, ,
    \label{eq:secIII-local-total-density-mapped}
\end{equation}
and resembles a nonlinear diffusion equation with cross-diffusion terms, which here result from concentration gradients and shape deformations of the line.

Now, since local reactive equilibria in mass-conserving reaction--diffusion systems serve as scaffolds for patterns~\cite{Halatek.Frey2018, Brauns.etal2020, Frey.Brauns2020}, one can approximate the membrane and cytosolic densities by their respective (stable) local reactive equilibria
\begin{equation}
    \left[\widetilde{m}(\mat,t),\tilde{c}(\mat,t)\right] 
    \rightarrow 
    \left[\widetilde{m}^\ast(\tilde{n}(\mat,t)),\tilde{c}^\ast(\tilde{n}(\mat,t))\right]
    \, .
\end{equation}
In doing so, one finds that only the metric $g$ and the (rescaled) local total density $\tilde{n}$ remain as the relevant dynamic variables in Eq.~\eqref{eq:secIII-local-total-density-mapped}. 
Evaluation of the spatial derivative inside the brackets in Eq.~\eqref{eq:secIII-local-total-density-mapped} results in:
\begin{multline}
    \mathcal{D}_t\tilde{n} \approx \partial_{\mat}\left[\frac{D_m \partial_{\tilde{n}}\widetilde{m}^\ast+D_c \partial_{\tilde{n}}\tilde{c}^\ast}{g} \, \partial_{\mat} \tilde{n} \right. \\
    \left.-\frac{1}{2}\frac{D_m \widetilde{m}^\ast+D_c\tilde{c}^\ast}{g^2} \, \partial_{\mat}g \right] \,.
    \label{eq:secIII-local-total-density-mapped-closed}
\end{multline}
The first term inside the brackets in Eq.~\eqref{eq:secIII-local-total-density-mapped-closed} describes diffusive mass-redistribution of $\tilde{n}$, and the second term contains higher-order nonlinear contributions to mass-redistribution originating from geometry deformations.
To proceed, we consider several limiting cases. 
If $g$ is independent of $\tilde{n}$, then Eq.~\eqref{eq:secIII-local-total-density-mapped-closed} is an equation with a linear feedback (first term, linear order in $\tilde{n}$) together with a forcing term (second term, zeroth order in $\tilde{n}$).
The stability of such equations is in general independent of the zeroth-order forcing term, and only depends on the linear-order feedback term.
In contrast, if the normal velocity of the curve is an arbitrary function of the local total density $\tilde{n}$, then according to Eq.~\eqref{eq:secIII-comoving-timederiv-metric} the dynamics of the metric is governed by $\mathcal{D}_t g(\mat,t)= -2\,g\,\kappa\,v_n(\tilde{n})$.
If we expand around an initially flat configuration of the line, $g = 1$ and $\kappa = 0$, then gradients of the metric will always vanish to linear order in time, so that the second term in Eq.~\eqref{eq:secIII-local-total-density-mapped-closed} still drops out.
In the most general case, the metric $g$ could be an arbitrary function of the local total density $\tilde{n}$.
In that case, we need to assume that deformations are weakly varying near onset of pattern formation such that ${\partial_{\mat}g \ll 1}$, which, again, implies that instabilities are dominated by the first term in Eq.~\eqref{eq:secIII-local-total-density-mapped-closed}. 
Without this approximation, no statement about instabilities can be made from Eq.~\eqref{eq:secIII-local-total-density-mapped-closed} and one would need to perform a (weakly) nonlinear analysis instead (assuming that the instability is supercritical)~\cite{Cross.Hohenberg1993}, which is a challenging task for our problem.
Using this approximation, we find that the stability of the system against spatial perturbations is determined by the effective diffusion coefficient in the first term of Eq.~\eqref{eq:secIII-local-total-density-mapped-closed}.
Since the metric $g$ is always positive by definition, the effective diffusion coefficient of $\tilde{n}$ becomes negative (leading to anti-diffusion) if ${D_m \partial_{\tilde{n}}\widetilde{m}^\ast+D_c \partial_{\tilde{n}}\tilde{c}^\ast < 0}$. 
Hence, a homogeneous steady state becomes unstable to spatial perturbations if:
\begin{equation}
   \frac{\partial_{\tilde{n}}\tilde{c}^\ast}{\partial_{\tilde{n}}\widetilde{m}^\ast}=\partial_{\widetilde{m}}\tilde{c}^\ast(\tilde{n}) < -\frac{D_m}{D_c},
   \label{eq:secIII-lateral-instability-criterion-mapped}
\end{equation}
which, analogously to fixed planar geometries~\cite{Brauns.etal2020}, shows that lateral instabilities occur if the slope of the reactive nullcline, $\partial_{\widetilde{m}}\tilde{c}^\ast(\tilde{n})$, is steeper than the ratio of diffusion coefficients on the membrane and in the cytosol, $-D_m/D_c$.

Our key finding here is that the generalized slope criterion, Eq.~\eqref{eq:secIII-lateral-instability-criterion-mapped}, depends explicitly on the local total densities $\tilde{n}$ as well as the shape of the reactive nullcline ${\tilde{f}(\mat,t)=0}$, which is in general space and time dependent.
This finding differs sharply from the case of a fixed planar geometry, where the same form of a slope criterion holds but with the shape of the nullcline fixed in both space and time, therefore leaving the local total density as the only control variable~\cite{Brauns.etal2020}.
The space and time dependency of the reactive nullcline results in inhomogeneities, which suggests that the lateral stability of the system may vary between spatial regions and may also evolve over time, depending explicitly on the time evolution of the line conformation. 
This is illustrated in Fig.~\ref{fig:nullclines-inhomogeneity}, where we show how spatial variations in the metric $\sqrt{g}$ (at a given point in time) induce spatial inhomogeneities of the nullcline shape, which results in lateral instabilities in those segments of the line where the criterion given by Eq.~\eqref{eq:secIII-lateral-instability-criterion-mapped} is fulfilled.

\section{Pattern formation on deforming manifolds}
\label{sec:regional_instability}

\subsection{External control of manifold conformations}
\label{subsec:external_deform}

\begin{figure*}[tbh]
\includegraphics[width=\textwidth]{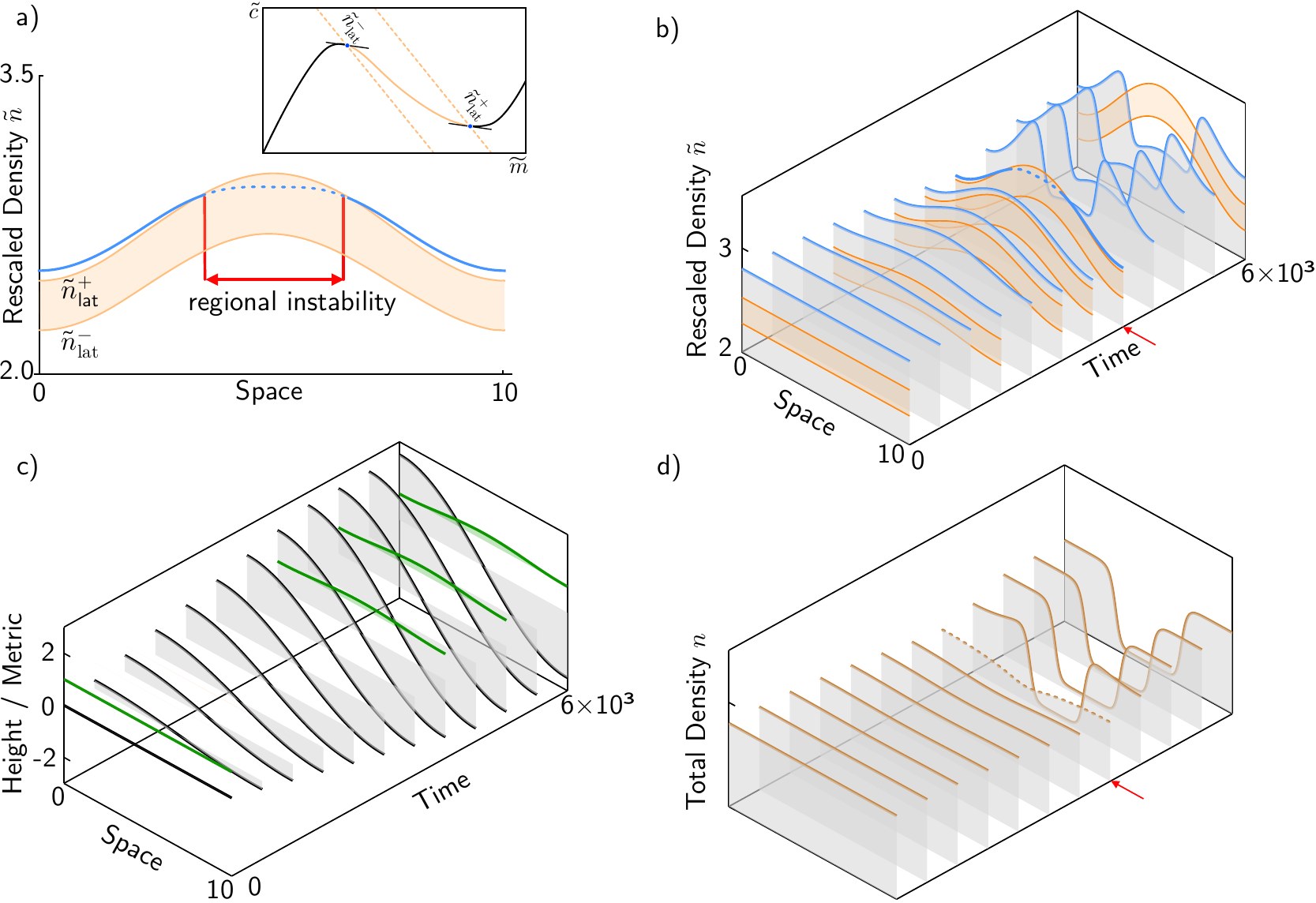}
\caption{%
Regional instability induced by line shape deformations. 
a) Snapshot of the rescaled local total density $\tilde{n}$ (blue solid line) and the region of lateral instability (orange area, the inset illustrates the definition of this area in phase space).
The snapshot further shows the onset of the regional instability as $\tilde{n}$ enters the orange area (dashed blue line).
b) Same as a), but for a set of time points (kymograph, where the grey shading visualizes the offset of the graph in the vertical direction).
A pattern forms when $\tilde{n}$ enters the region of instability; the red arrow indicates the onset of the instability and corresponds to the graph shown in a). 
For clarity, the region of instability is only shown for selected time points.
c) Time evolution of the line shape (black solid line) and metric (green solid line) as defined in Eq.~\eqref{eq:secIV-height-ext-deform-plateau}.
d) Actual local total density $n$; the red arrow indicates the onset of instability (same axes range as in b)).
}
\label{fig:ext_perturb_plataeu}
\end{figure*}

In the preceding sections, we established the theoretical framework to study reaction--diffusion systems on time-evolving one-dimensional manifolds.
So far, the analysis has been general, and we have not further specified exactly how the shape of the line deforms over time, or whether and how protein dynamics can feed back onto these deformations.

To illustrate the key points of pattern formation on lines that can change their conformation, we analyze below simple examples where we assume that the conformation of the line is controlled externally, i.e., we explicitly specify the temporal evolution of the line shape and investigate how the reaction--diffusion dynamics responds to these perturbations.
In detail, we consider the following scenario: we assume that the line shape is initially flat, and initialize the reaction--diffusion system such that the homogeneous steady state is stable against spatial perturbations (no pattern formation).
In other words, the total density is chosen such that the slope criterion, Eq.~\eqref{eq:secIII-lateral-instability-criterion-mapped}, is not fulfilled.
At time ${t=T_0}$, the shape of the line is then \textit{adiabatically} deformed from a straight conformation to a cosine-shape (Fig.~\ref{fig:ext_perturb_plataeu}c):
\begin{equation}
    h(x,t)
    =
    A(t)
    \cos\left(\frac{\pi x}{L}\right)
    \, ,
\label{eq:secIV-height-ext-deform-plateau}
\end{equation}
where the amplitude $A(t)$ is chosen to increase linearly from $0$ to $A_0$ during the time interval $[T_0,T_1]$ (ramp function):
\begin{align}
\label{eq:secIV-amplitude-ramp}
    A(t) 
    = 
    A_0 \color{gray}\times\color{black}
    \left\{ 
    \begin{array}{cc} 
        0, & \hspace{5mm} t<T_0 \\
        \frac{t-T_0}{T_1 - T_0}, & \hspace{5mm} T_0 \leq t \leq T_1 \\
        1, & \hspace{5mm} t > T_1 \\
        \end{array} 
    \right.
    \, .
\end{align}
The length of the time interval and the final amplitude $A_0$ are chosen such that the rate of line shape deformation is slow compared to the typical growth rate of unstable modes (small compared to ${\sim D_\text{c}q^2}$, where $q$ denotes the mode number; see Ref.~\cite{Brauns.etal2020} for details).
For clarity, we omit physical units in the following and specify typical length and time scales in an intracellular context in Appendix~\ref{appendix:parameters}.
Note that in our numerical analysis we explicitly use a Monge parameterization to describe the line conformations, and perturb the line shape by directly increasing the height $h(x,t)$ instead of imposing motion along the normal vector of the line.
This leads to a slight change in the equations, as the tangential velocity $v_\tau$ in Eq.~\eqref{eq:secIII-mapped-flux-membrane} can be omitted.\footnote{Consistent with our assumption of deformations along the normal direction, a tangential velocity only enters the equations if one chooses a parametrization in the Eulerian frame (as illustrated in Fig.~\ref{fig:parametrization}).
Here, to keep the analysis concise, we assume that the manifold is deformed along the vertical direction.
Hence, since the direction of the deformations coincides with the coordinate system in this case, the tangential velocity $v_\tau$ can be disregarded.}
To investigate the dynamics, we performed finite-element-method (FEM) simulations using the commercially available software \emph{COMSOL Multiphysics v5.6}.
The simulations show that the homogeneous steady state becomes laterally unstable for sufficiently large line shape deformations, and the concentration profile then gradually evolves into a mesa pattern along the spatial domain considered (Fig.~\ref{fig:ext_perturb_plataeu}b,c,d and Movie~1).
\begin{figure*}[tbh]
\includegraphics{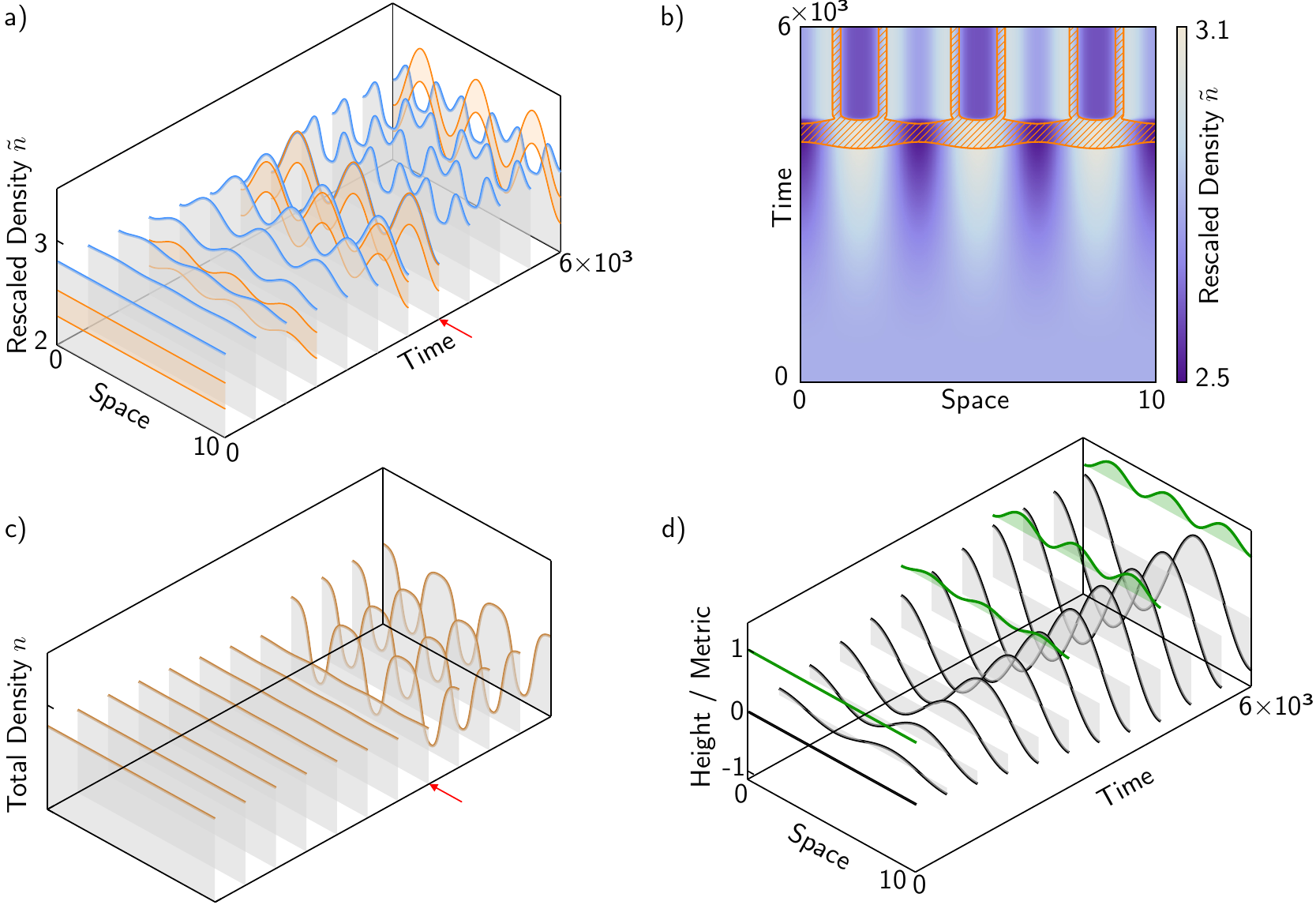}
\caption{%
The shape of the line acts as a template for patterns. 
a) A pattern-forming instability is triggered once the rescaled local total density $\tilde{n}$ (blue solid line) enters the region of instability (orange area).
For clarity, the region of instability is only shown for selected time points.
The red arrow indicates the onset of pattern formation.
b) Same kymograph as shown in a), but projected onto the space--time plane with the color-coding indicating the rescaled local total density $\tilde{n}$.
The orange hatched area indicates the values of $\tilde{n}$ which lie in the region of a lateral instability.
Note that a pattern forms once the orange hatched area appears (after a short lag time).
c) Actual local total density $n$ (same axes range as in a)).
Note that the rescaled density $\tilde{n}$ (shown in a) and b)) is non-uniform before the onset of a lateral instability (due to deformations), while the actual density initially remains nearly homogeneous and develops a spatial pattern once the system reaches the onset of instability.
d) This panel shows the line conformation (black solid line) and metric (green solid line), corresponding to Eq.~\eqref{eq:secIV-height-ext-deform-multiple-plateaus}.
Comparing with c) reveals that the high concentration regions of the pattern (plateaus) form precisely at the extrema of the height profile. 
}
\label{fig:ext_perturb_multiple_plateaus}
\end{figure*}

What is the mechanism underlying this lateral instability induced by the deformations that we impose on the line shape? 
To answer this question, we make use of the instability criterion for the rescaled densities, Eq.~\eqref{eq:secIII-lateral-instability-criterion-mapped}. 
Specifically, at each point in space and time, we determine the range of the (rescaled) total densities for which criterion~\eqref{eq:secIII-lateral-instability-criterion-mapped} is fulfilled.
To that end, we solve the equation ${\partial_{\widetilde{m}}\tilde{c}^\ast(\tilde{n}) = -D_m/D_c}$ for the rescaled total density $\tilde{n}$, and thereby determine an upper threshold $\tilde{n}_\text{lat}^+(x,t)$ and a lower threshold $\tilde{n}_\text{lat}^-(x,t)$, for which the instability criterion is fulfilled, i.e., for total densities that obey the inequality ${\tilde{n}_\text{lat}^- < \tilde{n} < \tilde{n}_\text{lat}^+}$.
Geometrically, these thresholds determine the points of the nullcline where the slope is equal to $-D_m/D_c$; see inset of Fig.~\ref{fig:ext_perturb_plataeu}a.

In contrast to the case of a fixed planar geometry, the existence and concentration range of a lateral instability here generally depends on space and time, since deformations cause spatial heterogeneities of the nullcline shape; see Eq.~\eqref{eq:secIII-mapped-reaction}.
This implies that shape deformations of the line may induce pattern-forming instabilities in regions of the line where the slope criterion~\eqref{eq:secIII-lateral-instability-criterion-mapped} is fulfilled (regional instabilities, see Fig.~\ref{fig:ext_perturb_plataeu}a).

Figure~\ref{fig:ext_perturb_plataeu}b shows the spatiotemporal dynamics of the rescaled local total density $\tilde{n}$ (blue solid line) and the region of lateral instability (orange shaded region).
For ${t<T_0}$, where the conformation of the line is flat, no pattern forms since $\tilde{n}$ lies outside the laterally unstable region.
Note that the region of lateral instability is spatially uniform for ${t<T_0}$ since one has a flat geometry with a metric ${g(x,t) = 1}$.
As the shape of the line is adiabatically deformed, the spatial profile of $\tilde{n}$ and the region of lateral instability also deform, eventually causing spatial sections of $\tilde{n}$ to enter the laterally unstable region. 
This event then induces a regional instability of the reaction--diffusion system and therefore leads to establishment of a pattern along the one-dimensional manifold considered (see Fig.~\ref{fig:ext_perturb_plataeu}a,b,d).

To conclude this section, we briefly summarize our key findings.
Essentially, the impact of line shape deformations is reflected in the spatially inhomogeneous nullcline shapes, which physically correspond to inhomogeneous reactive flows due to local length expansion and contraction and the corresponding changes in local particle concentrations.
This heterogeneity in the nullcline shape (dictated by the conformation of the line) leads to a non-uniform region of instability, which may induce (regional) pattern-forming instabilities in the reaction--diffusion dynamics as explained above.
We will in the following refer to this as a \textit{geometry-induced instability}, since changes in the shape of the manifold lead to spatial variations of the metric, which in turn affect the stability of the reaction--diffusion dynamics.
\begin{figure*}
\includegraphics{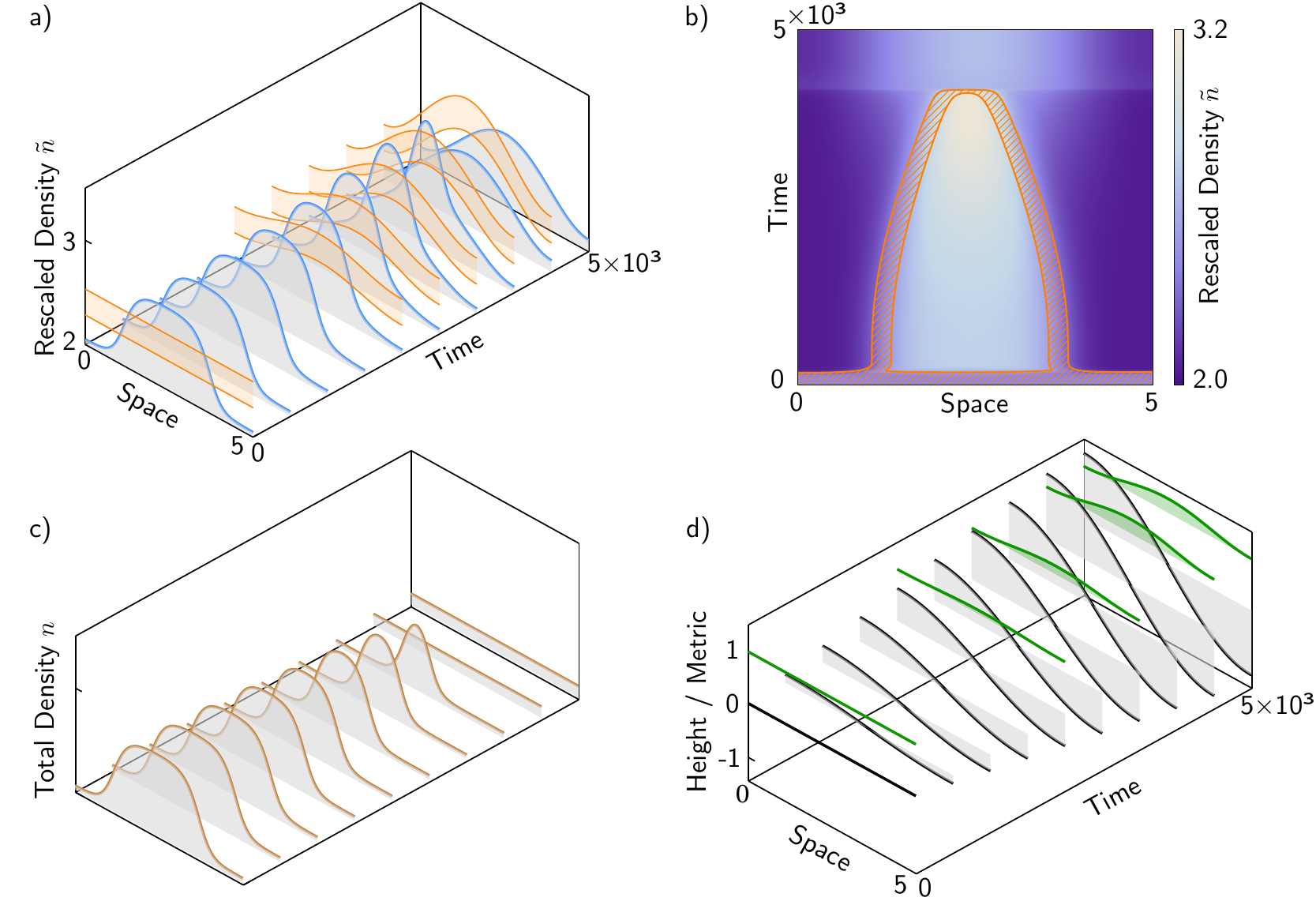}
\caption{
Shape deformations shift the interface and suppress patterns.
a) The total average density is chosen such, that initially a mesa pattern forms.
Due to deformations of the line shape, the interface (connecting the lower and upper plateau) is shifted such that the plateau width decreases (see discussion in main text).
For sufficiently large deformations, only a small region of the rescaled local total density $\tilde{n}$ (blue solid line) lies in the region of instability (orange area), which then leads to suppression of the pattern (see also c)).
b) Same kymograph as shown in a), but projected onto the space--time plane with the color-coding indicating the rescaled local total density $\tilde{n}$.
The orange hatched area indicates the values of $\tilde{n}$ which lie in the region of a lateral instability.
Note that the interface is first shifted and then suppressed as the plateau width becomes too small.
c) Actual local total density $n$ (same axes range as in a)).
d) Time evolution of the line shape (black solid line) and metric (green solid line), as defined in Eq.~\eqref{eq:secIV-height-ext-deform-shift}.
}
\label{fig:ext_perturb_shift}
\end{figure*}

\subsection{Shape deformations act as a template for patterns}
\label{subsec:geometry_template}

Above, we have shown that an externally controlled deformation of the line shape can induce lateral instabilities and thus lead to pattern formation.
While we have chosen a specific way to gradually deform the geometry, the principle that we have found is general: the high-concentration plateaus of the emerging pattern form at characteristic locations along the line determined by geometric features, here extrema of the height profile; see Fig.~\ref{fig:ext_perturb_plataeu}c and Fig.~\ref{fig:ext_perturb_plataeu}d.
Thus, the shape of the line acts as a kind of template for the patterns, with the low-concentration plateaus emerging in regions where the height profile has maximal slope.
In these regions, the metric of the line is at its largest, thus maximizing the local depletion of particles due to dilation in the local line geometry. 
This is a generic feature of the system, as we will explain in the following.

To elucidate this templating effect further, we repeat the analysis from the previous section, but this time consider a higher-harmonic shape deformation (Fig.~\ref{fig:ext_perturb_multiple_plateaus}c):
\begin{equation}
    h(x,t)=A(t)\cos\left(\frac{3\pi x}{L}\right),
    \label{eq:secIV-height-ext-deform-multiple-plateaus}
\end{equation}
where the amplitude $A(t)$ is defined by Eq.~\eqref{eq:secIV-amplitude-ramp} and the initial conditions are again chosen such that the homogeneous steady state is laterally stable.
In agreement with our previous results, a pattern-forming instability is triggered as soon as the rescaled total density $\tilde{n}$ enters the region of lateral instability (see Fig.~\ref{fig:ext_perturb_multiple_plateaus}a,b,c and Movie~2).
As we have chosen a higher-harmonic shape deformation for the line, multiple plateaus now form.
As expected, the high-density plateaus are located at the extrema of the line height, while the low-density plateaus are located in regions where the height profile of the line is steepest; compare Fig.~\ref{fig:ext_perturb_multiple_plateaus}c and Fig.~\ref{fig:ext_perturb_multiple_plateaus}d.

This is effect is due to the non-uniform deformation of the line shape.
In particular, while the total length of the height profile increases over time, the local length of individual line segments barely changes at the extrema of the height profile.
Thus, the growth of the line's contour length occurs primarily in regions with a steep slope of the height profile, where the metric is largest.
In these regions, where ${\partial_t \sqrt{g(x,t)} > 0}$, the concentration of particles will be diluted.
Thus, even when starting from an initially homogeneous concentration profile, such geometric effects alone lead to a redistribution of particles and density inhomogeneities $n(x,t)$ along the line (Fig.~\ref{fig:ext_perturb_multiple_plateaus}c).
If this effect is coupled to the onset of a lateral instability, then troughs in the density profile will further decrease, while hills (located at the extrema of the height profile where ${\sqrt{g(x,t)} \approx 1}$) will increase (Fig.~\ref{fig:ext_perturb_multiple_plateaus}a,b,d).
In other words, the mass-redistribution instability~\cite{Brauns.etal2020} will further amplify geometry-induced density inhomogeneities.

\subsection{Interface shift and pattern suppression}
\label{subsec:pattern-interface-shift}

So far, we have investigated cases where the homogeneous steady state was laterally stable for a planar geometry, and we induced an instability only by deforming the line shape.
We now ask how shape deformations affect already established patterns.
To address this question, we choose the initial total density such that the homogeneous steady state is laterally unstable, thereby resulting in the formation of a mesa pattern (Fig.~\ref{fig:ext_perturb_shift}a,c and Movie~3).

We initialize the reaction--diffusion system and its steady-state pattern on an initially planar geometry (straight line).
Then, following the same procedure as before, we deform the line shape adiabatically (Fig.~\ref{fig:ext_perturb_shift}d): 
\begin{equation}
    h(x,t)=A(t)\cos\left(\frac{\pi x}{L}\right),
    \label{eq:secIV-height-ext-deform-shift}
\end{equation}
where the amplitude $A(t)$ is defined by Eq.~\eqref{eq:secIV-amplitude-ramp}.
We find that the deformation in the line's shape (Fig.~\ref{fig:ext_perturb_shift}d) gradually changes the pattern profile from a mesa to a narrow peak pattern, until eventually the peak disappears altogether (Fig.~\ref{fig:ext_perturb_shift}c).
There are two major underlying reasons for these observations: 
First, for the two-component reaction--diffusion system that we study here, it has been shown that the interface position (which connects the lower and upper plateau of a mesa pattern) depends only on the average total density $\langle n \rangle$, as long as the system size is larger than the typical length scale of the pattern interface~\cite{Brauns.etal2020}.
Thus, altering the average total density will shift the interface as a consequence of mass-conservation, where addition or removal of mass leads to a respective increase or decrease of the plateau width.
Since deformations of the line shape effectively lead to a  depletion of the total average density due to an increase in the total length of the spatial domain, this explains the shift of the pattern interface.
Second, for large enough shape deformations, only a small part of the  system lies in a region of lateral instability; see Fig.~\ref{fig:ext_perturb_shift}a and Fig.~\ref{fig:ext_perturb_shift}b. 
Once the size of this region becomes comparable or even smaller than the typical length scale of the interface, pattern formation is suppressed.

\subsection{Self-organized mechanochemical coupling}
\label{subsec:dynamic_coupling}

In the previous sections, we have gained basic insights into how shape deformations affect the pattern formation of mass-conserving reaction--diffusion systems.
We now consider a more intricate scenario where the dynamics of the conformation of the line is explicitly coupled to the density fields on the one-dimensional manifold.
In other words, we incorporate a feedback loop between the line shape and the reaction--diffusion dynamics of the density fields.
In general, there are many ways to implement such a coupling. 
For example, the local concentration of proteins can drive shape deformations as well as protein transport on and onto the manifold through local bending of the membrane~\cite{Goychuk.2019, Yuan.etal2021, Mahapatra.etal2021}, active stresses in the form of myosin contractility~\cite{Shlomovitz.Gov2007, Mietke.etal2019a, Mietke.etal2019b} or actin polymerization~\cite{Gov.Gopinathan2006, Shlomovitz.Gov2007}.

\begin{figure}[t]
\includegraphics{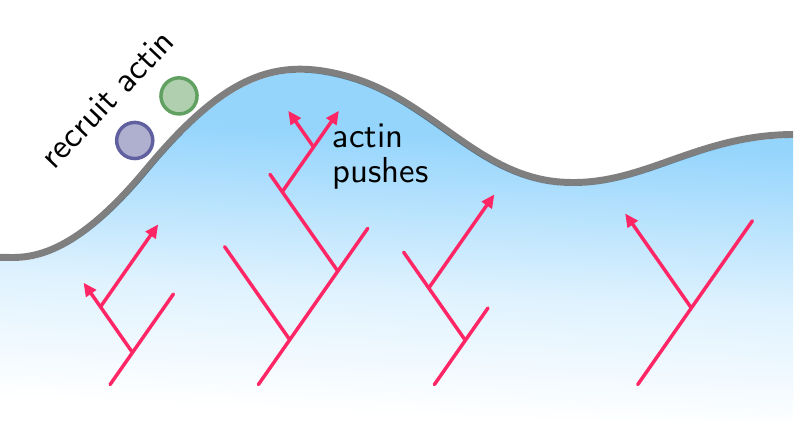}
\caption{%
Illustration of actin-driven membrane deformations.
Membrane-bound and cytosolic proteins enhance the local polymerization of actin filaments, which in turn push the membrane outwards.
These effects can most simply be modelled by an isotropic cytosolic pressure that drives membrane motion along its normal vector (see Eq.~\eqref{eq:secIV-normal_vel_density}).
}
\label{fig:recruitment-actin}
\end{figure}

Furthermore, there are many ways to account for the mechanical properties of an elastic (or viscoelastic) manifold. 
For example, consider a cell membrane; the line we have considered so far can be seen as a one-dimensional projection of such a membrane. 
The conformation of a membrane is characterized by an elastic energy that generally contains both a bending energy term and a surface tension term. 
Which of these contributions dominates depends on the system in question~\cite{Seifert.1997, Lee.etal2021, Lipowsky.2022}. 
Here, we study a conceptually simple example where we assume that the line can be regarded as a fluid-like substrate, i.e., we disregard mechanical properties of the line such as bending rigidity.
To further simplify our system, we additionally assume that the tension $\gamma$, cf. Eq.~\eqref{eq:secIV-normal_vel_density}, is spatially uniform.
The feedback loop between the particle density and the membrane conformation is implemented by assuming that shape deformations are locally driven by the local total density of proteins.
Specifically, we consider the following form for the normal velocity $v_n$:
\begin{equation}
    v_n
    =
    \mu\,
    \big[
    m(\eul,t)+c(\eul,t)
    \bigl] 
    + \gamma \, \kappa(\eul,t) \,.
    \label{eq:secIV-normal_vel_density}
\end{equation}
The parameter $\mu$ denotes the coupling strength between the local protein density and the normal velocity of the line.
Physically, one may interpret this term as a protein-controlled recruitment and polymerization rate of actin filaments that drive outwards motion of the membrane through an effective pressure~\cite{Gov.Gopinathan2006, Shlomovitz.Gov2007, Mogilner.2009} (see Fig.~\ref{fig:recruitment-actin} for an illustration).
The second term accounts for the Laplace pressure caused by surface tension effects due to the local curvature $\kappa(\eul,t)$.
Phenomenologically, the first term thus describes local growth of the membrane that is proportional to the local total density of proteins, while the second term counteracts this effect by minimizing the membrane area (which is a length in our case since we consider a one-dimensional projection of the membrane surface).

In the Monge representation of the line, and mapping to rescaled variables $\widetilde{m}$ and $\tilde{c}$ (Sec.~\ref{secIII-two-comp-deform-geom}), one can rewrite Eq.~\eqref{eq:secIV-normal_vel_density} as:
\begin{equation}
    v_n
    =
    \mu \, 
    \frac{1}{\sqrt{g}}
    \bigl[
    \widetilde{m}(x,t)+\tilde{c}(x,t)
    \big]
    +
    \gamma \, 
    \frac{1}{g^{3/2}} \, 
    \partial_x^2 h(x,t) 
    \,.
\label{eq:secIV-normal_vel_density_mapping}
\end{equation}
The time evolution of the line conformation is then obtained from Eqs.~\eqref{eq:secIII-mapped-system-c} and~\eqref{eq:secIV-normal_vel_density_mapping}:
\begin{align}
    \partial_th(x,t)
    &=
    \sqrt{g(x,t)} \, 
    v_n
    \nonumber \\
    &=
    \mu\,
    \big[ 
    \widetilde{m}(x,t)+\tilde{c}(x,t)
    \big]
    +
    \gamma\, 
    \frac{1}{g} \, 
    \partial_x^2 h(x,t) 
    \,.
\label{eq:secIV-evol_eq_height}
\end{align}
Eq.~\eqref{eq:secIV-evol_eq_height} for the height profile of the one-dimensional manifold, together with Eqs.~\eqref{eq:secIII-mapped-system-a}~and~\eqref{eq:secIII-mapped-system-b} for the density profiles of the proteins, provide a closed set of equations governing the self-organized dynamics of a two-component MCRD system on a deforming manifold with mechanochemical coupling.

In general, the dynamics of the mechanochemically coupled system is expected to depend on the relative time scales of line shape deformation and protein pattern formation.
The former is determined by the coupling strength $\mu$ and the total average density $\langle \tilde{n} \rangle$, and yields the characteristic time scale for the growth of the line's length, ${t_\text{G} = L_0/ ( \mu\,\langle \tilde{n} \rangle )}$.
The latter is dominated by cytosolic redistribution of particles and thus provides a typical time scale of diffusion ${t_\text{D}=L_0^2/D_c}$.
Hence, one may define a dimensionless number that relates the time scales of diffusion and growth, which we call in analogy to fluid dynamics the Péclet number
\begin{equation}
  \text{Pe} 
  \coloneqq 
  \mu \langle \tilde{n}\rangle L_0/D_c
  \, .
\label{eq:peclet}
\end{equation}
For small values of the Péclet number, ${\text{Pe} \ll 1}$, diffusion is much faster than the dynamics of line shape deformations.
In particular, in the limiting case of ${\text{Pe} \rightarrow 0}$, the dynamics of shape deformations becomes infinitely slow on the time scales of diffusive mass redistribution.
This is equivalent to abolishing mechanochemical coupling altogether, $\mu \rightarrow 0$, where the protein patterns approach a stationary state.
We provide a systematic analysis of the impact of this parameter on the pattern-forming dynamics in Sec.~\ref{subsec:time_scales}.

In the following, we will first explore the system's dynamics through FEM simulations, and find a broad variety of dynamic patterns.

\subsubsection{Oscillations}
\label{subsubsec:oscillations}

First, we performed simulations of small confined systems with reflecting boundaries. 
The initial total average density was chosen such that the reaction--diffusion system is laterally unstable and therefore generates a mesa pattern.
For the initial conformation of the line, we selected a flat state with $h(x,0)=0$.

\begin{figure}[!t]
\includegraphics{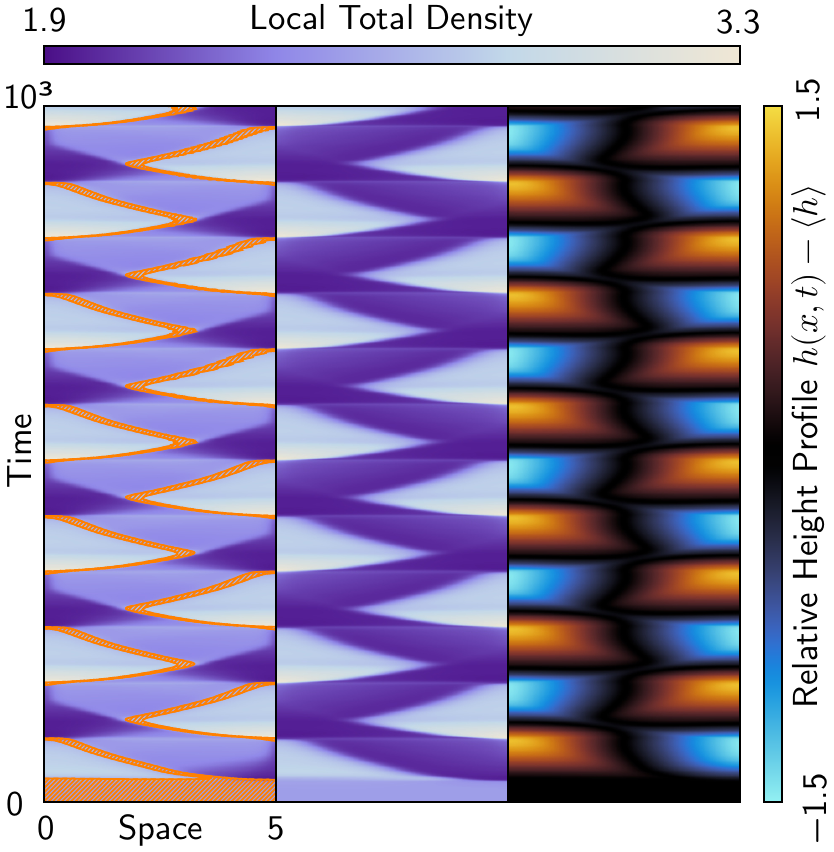}
\caption{%
Coupling the reaction--diffusion dynamics to the shape of the line leads to self-organized spatiotemporal patterns. 
The left panel shows a kymograph of the spatiotemporal dynamics of the rescaled total density $\tilde{n}$ and illustrates the emergence of oscillations.
The orange shaded area corresponds to values of $\tilde{n}$ which fulfill the instability criterion Eq.~\eqref{eq:secIII-lateral-instability-criterion-mapped}.
Note that, as the pattern amplitude on one boundary of the domain disappears (due to deformations of the line shape, see right panel), a regional instability is induced at the opposite boundary.
This interplay between reaction--diffusion dynamics and line shape deformations drives the spatiotemporal dynamics.
The middle panel shows the actual local total density $n$.
The right panel illustrates the patterns in the relative height profile, defined as ${h(x,t)-\langle h \rangle}$,  where the average height $\langle h \rangle$ is proportional to the average total density and time ${\langle h \rangle \sim \langle \tilde{n} \rangle \, t}$.
}
\label{fig:geometry_coupling_osc_confined}
\end{figure}

Interestingly, although the two-component reaction--diffusion system shows only stationary patterns for a static line shape, we here find self-organized oscillations. 
These spatiotemporal patterns must therefore clearly be due to the mechanochemical feedback between the protein pattern and the line shape  (Fig.~\ref{fig:geometry_coupling_osc_confined}, Movie~4).
However, the question remains as to how the mechanism driving these oscillations relates to the geometry effects discussed earlier (Secs.~\ref{subsec:external_deform}--\ref{subsec:pattern-interface-shift}).

\begin{figure}[t]
\includegraphics{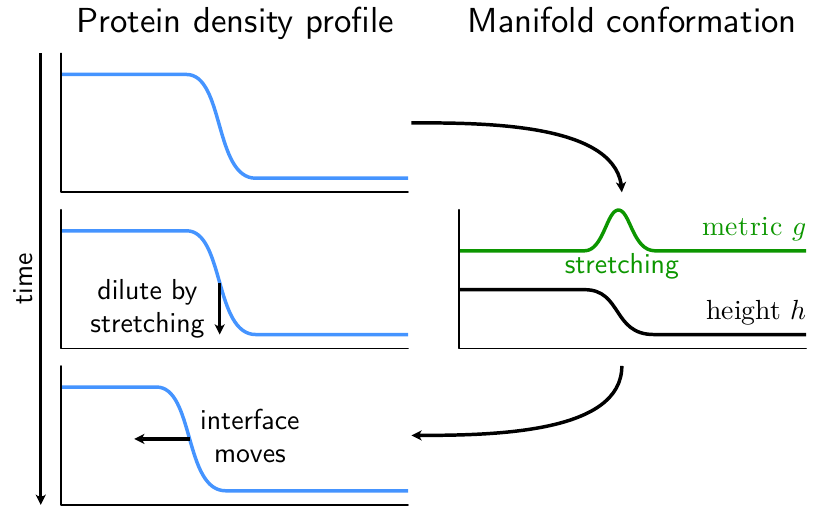}
\caption{%
Illustration of the mechanism that drives the motion of the protein pattern's interface.
The one-dimensional manifold moves along its normal vector, with a velocity that is proportional to the local protein density.
Therefore, gradients in protein density lead to gradients in the height profile, thus stretching the line at the location of the protein pattern's interface.
Since local stretching of the line corresponds to local dilution of the protein density, the interface of the protein pattern moves.
}
\label{fig:interface_motion_mechanism}
\end{figure}

From the first term in the equation for the normal velocity $v_n$, Eq.~\eqref{eq:secIV-normal_vel_density}, we deduce that an initial pattern formed on a flat line destabilizes this line conformation by inducing faster growth of the height profile at the pattern's peaks than at its valleys.
The resulting change in the line shape geometrically corresponds to local length dilations and contractions. 
Here, where we started from a flat line conformation, the slope of the height profile and thus the local contour length of the line grows fastest at the pattern's interfaces between two plateaus. 
To illustrate how these changes in the line's geometry affect the dynamics of the protein pattern, let us for now suppose that we initiate the system with a protein pattern consisting of two plateaus (mesa pattern) and a flat conformation of the line (see Fig.~\ref{fig:interface_motion_mechanism}).
Then, the growth of the local line length at the interface (connecting the lower and upper plateau) will locally dilute the density of proteins.
As a consequence, the lower plateau will expand at the expense of the upper plateau, pushing the interface towards the upper plateau (see Fig.~\ref{fig:interface_motion_mechanism}).
Since line shape deformations also alter the region of instability of the reaction--diffusion system, large enough deformations will suppress the initial pattern and trigger a regional instability at the opposite side of the geometry (see Fig.~\ref{fig:geometry_coupling_osc_confined}).
As the upper plateau grows at the opposing side, it gradually restores the height profile to a flat conformation.
After the plateau pattern is fully re-established at the opposing side and the conformation has returned to a flat conformation, the cycle repeats.
This intricate interplay between the dynamics of the line shape and the reaction--diffusion system (mechanochemical feedback) is the key mechanism that leads to spatiotemporal oscillations.

\subsubsection{Traveling waves}
\label{subsubsec:traveling_waves}

For spatial domains much larger than the wavelength $\lambda_c$ of the fastest growing mode in the dispersion relation (Appendix~\ref{appendix:lsa_2comp}), the two-component mass-conserving reaction--diffusion system initially leads to the formation of patterns consisting of multiple plateaus or peaks.\footnote{For the specific system that we consider here, these are always mesa patterns (see Appendix~\ref{appendix:reaction_term}).}
The wavelength of this initial pattern is well approximated by $\lambda_c$.
However, this initial pattern is not stable and slowly coarsens to a single peak or interface~\cite{Arkin.etal2007,  Ishihara.etal2007, Nie.etal2018, Brauns.etal2021, Subramanian.etal2021, Weyer.etal2022}.

Let us now again consider how the dynamics is changed when there is a mechanochemical coupling between the density profile emerging from the reaction--diffusion system and the conformation of the line.
As in Sec.~\ref{subsubsec:oscillations}, we choose a flat height profile ${h(x,0) = 0}$ for the initial line conformation, and a homogeneous concentration of proteins with a slight random perturbation around this state; however, we now impose periodic boundary conditions for both, the reaction--diffusion dynamics and the line's shape.

\begin{figure}[t] 
\includegraphics{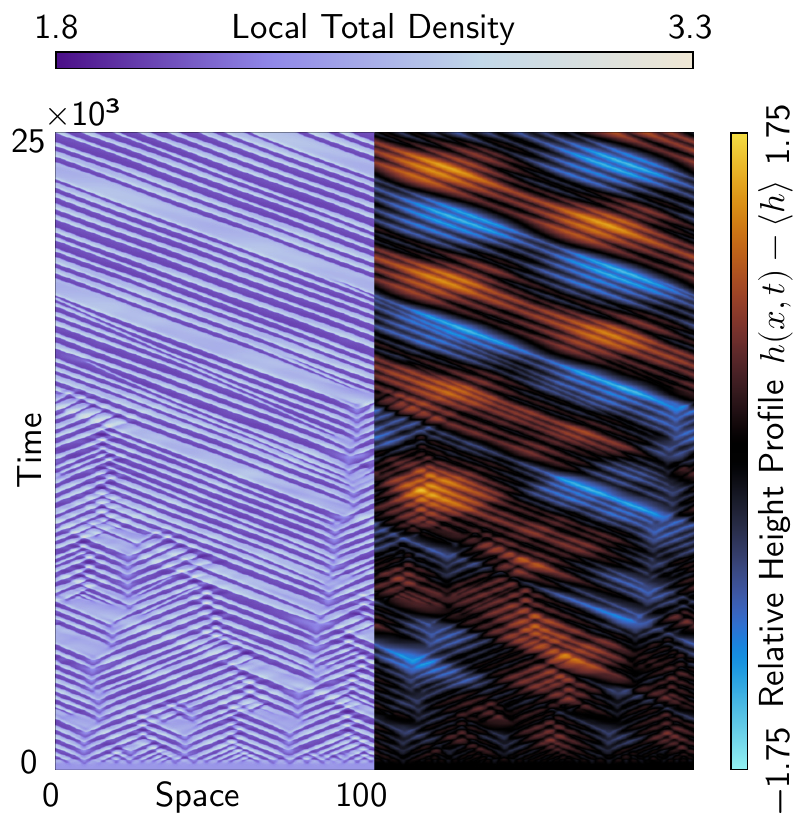}
\caption{%
Emergence of traveling wave patterns for Péclet numbers ${\text{Pe} < 0.6}$.
The left panel shows a kymograph of the spatio\-temporal dynamics of the actual local total density $n$, and the right panel depicts the relative height difference ${h(x,t)-\langle h \rangle}$.
Wave fronts emerge and vanish at specific points along the spatial domain considered. 
The positions of these events depend on the initial condition, which in our FEM simulations is a small random perturbation around the homogeneous steady state.
For long times, the system self-organizes into periodic traveling wave fronts.
}
\label{fig:geometry_coupling_traveling_waves}
\end{figure}

In our FEM simulations, we observe that propagating density waves and accompanying waves in the height profile arise at specific points in space (``sources'')  (Fig.~\ref{fig:geometry_coupling_traveling_waves}, Movie~5).
Each of these sources gives rise to two waves that travel in opposite directions and, given the periodic boundary conditions, mutually annihilate at specific points  on the spatial domain considered (``sinks'').
The position of these sources and sinks depend on the initial conditions, that is, the slight perturbations of the initially homogeneous density profiles.
Furthermore, we observe that these sources and sinks slowly migrate in space, and eventually meet and annihilate for large times.
The steady-state pattern then consists of periodic traveling wave fronts, as shown in Fig.~\ref{fig:geometry_coupling_traveling_waves}.
Importantly, we note that the system selects a typical wavelength for large times.
Hence, our results suggest that the coarsening process is interrupted if the dynamics is explicitly coupled to the geometry. 

\begin{figure*}[t]
\includegraphics{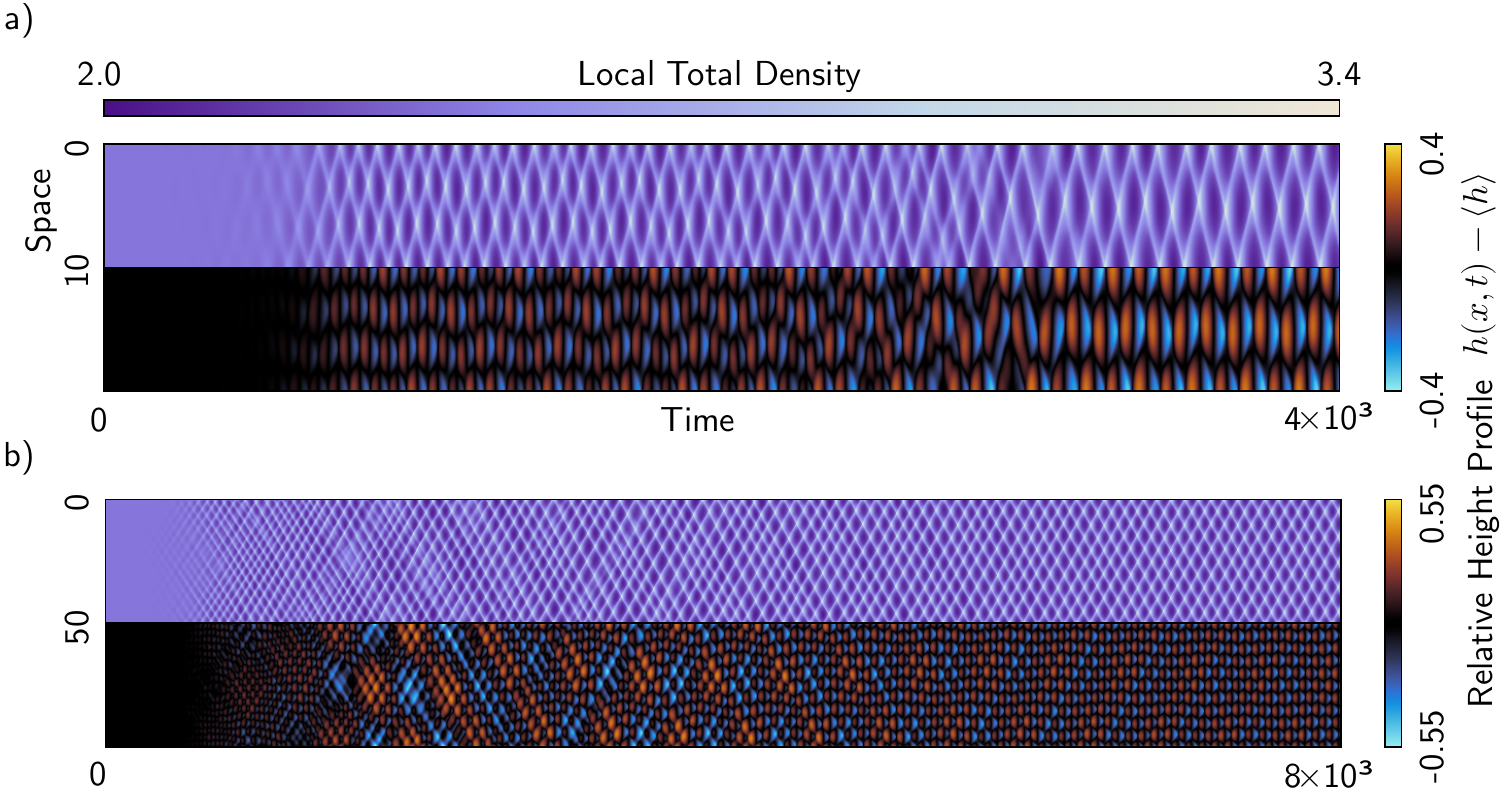}
\caption{%
Emergence of standing wave patterns for ${\text{Pe} > 0.6}$.
a) Spatiotemporal dynamics of the actual local total density $n$ and the relative height difference ${h(x,t)- \langle h \rangle}$ for system size ${L_0 = 10}$.
As for traveling waves, the dynamics settles on a specific wavelength over long times.
b) Same FEM simulation and parameters as shown in a), but for a larger system size ${L_0 = 50}$.
The final wavelength at long times is identical to that shown in a). 
}
\label{fig:geometry_coupling_standing_waves}
\end{figure*}

\subsubsection{Standing waves}
\label{subsubsec:standing_waves}

Depending on the relative magnitude of the characteristic time scales of changes in the line shape and the mass redistribution---the Péclet number Eq.~\eqref{eq:peclet}---we observe a transition from traveling wave patterns to standing wave patterns (Fig.~\ref{fig:geometry_coupling_standing_waves} and Movie~6).

For the parameter combination used in this study (Appendix~\ref{appendix:parameters}), the transition from traveling waves to standing waves occurs at a critical value of $\text{Pe}_c \simeq \SI{0.6}{}$.
This suggests that standing wave patterns emerge if the time scales of line shape dynamics and diffusive redistribution of proteins are comparable, whereas the emergence of traveling wave patterns requires that diffusive redistribution of proteins is the dominant (fastest) time scale.
Moreover, as for the traveling waves in Sec.~\ref{subsubsec:traveling_waves}, we find that the system selects a typical wavelength for large times (see Fig.~\ref{fig:geometry_coupling_standing_waves}a,b). 
Notably, for large domains, we observe that the pattern wavelength at small times is larger than the final wavelength in steady state (see Fig.~\ref{fig:geometry_coupling_standing_waves}b).
This again indicates that coarsening in the system seems to be interrupted.

\subsection{Tuning the relative time scales of the conformational dynamics and diffusive transport}
\label{subsec:time_scales}

In the previous sections, we found that coupling a MCRD system on a one-dimensional manifold with deformations of this manifold can lead to rich spatiotemporal dynamics. 
In FEM simulations of the coupled system we have observed oscillations or traveling waves, even though the MCRD system on a static manifold would typically approach a stationary steady state through coarsening.
As discussed above, such a mechanochemical coupling introduces an additional time scale that competes with the typical time that the MCRD system requires to generate a protein pattern, see Eq.~\eqref{eq:peclet}.

But how in detail does the dynamics of shape deformations affect the formation of protein patterns through reactions and diffusion?
Here, we answer this question by performing numerical parameter sweeps.
For convenience, we first introduce dimensionless quantities by rescaling spatial coordinates and time, 
\begin{subequations}
\begin{equation}
\begin{split}
    \{x, \, h\} \rightarrow L_0 \color{gray}\times\color{black} \{x',\,h' \}, \, \text{and}\quad
    t \rightarrow D_c^{-1} \, L_0^2 \color{gray}\times\color{black}  t' \, ,
\end{split}
\end{equation}
and thus also velocities, $v \rightarrow D_c \, L_0^{-1} \color{gray}\times\color{black} v'$.
Furthermore, we also rescale particle densities,
\begin{equation}
    \{\tilde{n}, \, \tilde{c}, \, K_d \} \rightarrow L_0^{-1} \color{gray}\times\color{black} \{\tilde{n}', \, \tilde{c}', \, K_d' \} \, ,
\end{equation}
and all control parameters:
\begin{align}
    \{D_m, \, \mu, \, \gamma \} &\rightarrow D_c \color{gray}\times\color{black} \{D, \, \mu', \, \gamma' \}\, , \\
    \{k_\text{on}, \, k_\text{fb}, \, k_\text{off} \} &\rightarrow D_c \, L_0^{-2} \color{gray}\times\color{black} \{k_\text{on}', \, k_\text{fb}', \, k_\text{off}' \} \, .
\end{align}
\end{subequations}
Here, we have grouped parameters with identical units of measurement and indicate their non-dimensionalized counterparts by the prime symbols on the right-hand side.
The non-dimensionalized equations are shown in Appendix~\ref{appendix:nondimensionalized_equations}, where we have dropped the primes to simplify notation.
Due to the non-dimensionalization, all variables are scaled to the system size.
Here ${D = D_m / D_c}$ denotes the ratio of diffusion constants, and the Péclet number relating the time scale of shape dynamics to the time scale of (cytosolic) diffusion is now given by ${\text{Pe} = \mu' \langle n'\rangle}$. 

\begin{figure}[!t]
\includegraphics{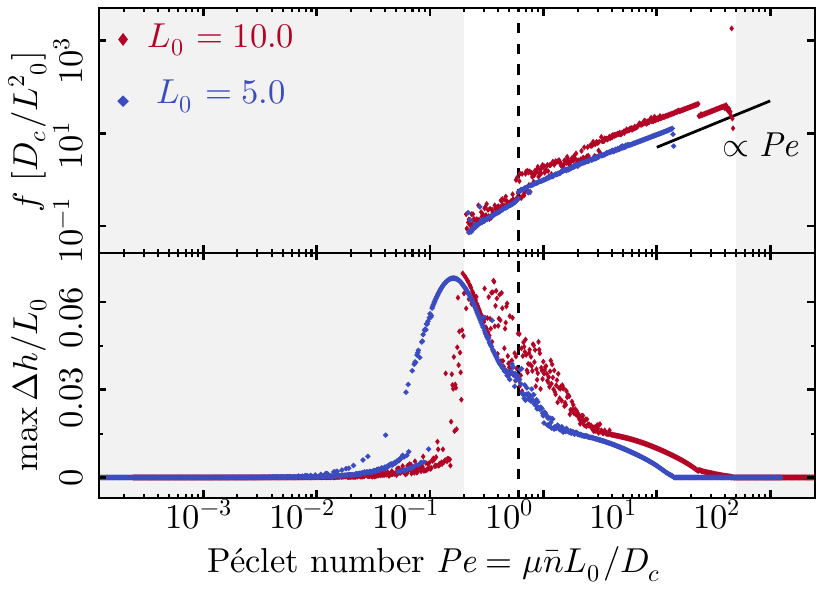}
\caption{%
Features of the oscillatory dynamics of the one-dimensional manifold in our simulations, as a function of the Péclet number.
In our FEM simulations, we monitor the height difference, ${h(L, t) - h(0, t)}$, of the one-dimensional manifold as a function of time.
We then determine the frequency of the oscillations (top) and the maximal amplitude of the oscillations (bottom).
In the limit of small Péclet number, the height profile remains static, thus leading to stationary patterns that persist until a critical value of $\text{Pe} \sim 0.2$.
For sufficiently large Péclet number, we observe an onset of oscillations, whose frequency increases (approximately) linearly.
When the Péclet number exceeds a second critical value of ${\mu/D_c \simeq 50}$, all dynamics vanishes.
}
\label{fig:time_scales}
\end{figure}

From our numerical parameter study, we find that the system can remain in a stationary state as long as the Péclet number is sufficiently small, below a finite critical value of ${\text{Pe} \lesssim \SI{0.2}{}}$.
However, if we increase the Péclet number beyond this value, then we find a discontinuous onset of oscillations (Fig.~\ref{fig:time_scales}).
When further increasing the Péclet number, the oscillation frequency increases.
When the line moves at a high velocity (corresponding to large oscillation frequencies), the coupling to the manifold quickly redistributes proteins and thus flattens out protein density gradients via an effective artificial diffusion. 
For example, regions with a high concentration of proteins grow faster, hence reducing the local concentration of proteins by virtue of mass conservation.
Thus, above a large Péclet number of ${\text{Pe} \gtrsim \SI{50}{}}$, we find a suppression of both protein pattern formation and consequently oscillations (Fig.~\ref{fig:time_scales}).
Interestingly, we find a doubling of the measured oscillation frequencies at intermediate Péclet numbers ${\text{Pe} \sim \SI{0.6}{}}$, which for large system sizes corresponds to the onset of standing wave patterns, as discussed in the previous section.

To conclude, through our numerical parameter study we have learned how the shape dynamics affects protein pattern formation on the one-dimensional manifold.
The qualitative dynamics of protein pattern formation remains largely unaffected by the deformations as long as the Péclet number, which relates the time scale of the line's deformations to the time scale of protein diffusion, is sufficiently small.
However, there are two qualitative changes with increasing Péclet number:
First, one observes a discontinuous onset of oscillatory patterns, and second, at high Péclet numbers all patterns are gradually extinguished.

\section{Discussion}
\label{sec:discussion}

We investigated the dynamics of a two-component mass-conserving reaction--diffusion system on a dynamically deforming one-dimensional manifold embedded in two-dimensional space.
To shed light on how deformations of the line influence pattern formation, we first studied a scenario where these shape deformations are externally controlled and occur on a time scale much larger than that of the intrinsic dynamics of the reaction--diffusion system (adiabatic deformations).
Next, we considered a feedback loop between shape deformations and the reaction--diffusion dynamics.
To keep the analysis simple and concise, we assumed a fluid-like substrate with a growth rate proportional to the local total protein density.
We found that shape deformations induce spatially non-uniform pattern-forming instabilities, which we refer to as \emph{regional instability}.
Moreover, our analysis shows that the shape dynamics may also (regionally) suppress protein patterns and spatially shift already established protein pattern interfaces.
Despite its simplicity, the model already shows a surprisingly wide range of dynamic patterns, such as oscillations and traveling waves. They emerge as a direct consequence of the interplay between shape deformations and reaction--diffusion dynamics.

Based on the local equilibria theory~\cite{Brauns.etal2020}, 
we then derived a criterion that links the onset of instabilities to the slope of the reactive nullcline in phase-space.
Specifically, we find that the nullcline shape becomes spatially non-uniform because the metric of the geometry enters the dynamics.
This differs sharply from the case of a flat static geometry~\cite{Brauns.etal2020}, where the nullcline shape is uniform in space and time.
Our analysis shows that the interplay between the dynamics of the local total density, an important control parameter for pattern formation in mass-conserving reaction--diffusion systems~\cite{Halatek.Frey2018,Brauns.etal2020,Frey.Brauns2020}, and the metric is key to understanding the phenomenology of the system.
From a physical point of view, the underlying mechanism of the observed dynamics lies in the local dilution and enrichment of particle densities by local length contraction and extension, respectively, which occur concomitantly with dynamic changes in the shape of the line.

We further showed that the existence of dynamic patterns crucially depends on the characteristic time scales of shape deformations and (cytosolic) diffusive mass redistribution, which we quantified by defining a (dimensionless) Péclet number that describes the ratio between these two time scales. 
Depending on the Péclet number, we identified two distinct asymptotic limits:
\begin{enumerate*}[label=(\roman*)]
    \item for small values of the Péclet number, one finds quasi-stationary patterns as the line deforms on a time scale much larger than the typical time scale of mass redistribution (diffusion-dominated regime), and
    \item for large values of the Péclet number, pattern formation is suppressed due to an instantaneous and large deformation rate of the line, which prevents the establishment of gradients in the particle densities (growth-dominated regime).
\end{enumerate*}
Between these two limiting regimes, the mechanochemical coupling yields a rich dynamics including oscillations and traveling wave patterns.

We found that the impact of deformations of the manifold is specified by a simple criterion, Eq.~\eqref{eq:secIII-lateral-instability-criterion-mapped}, which predicts the onset of regional instabilities.
Strikingly, the only geometric information that enters this criterion is the metric of the manifold.
This implies that the instability criterion is generic for mass-conserving reaction-diffusion systems, regardless of the exact cause and the associated mechanical forces that lead to shape deformations.
Here, we assumed a fluid-like manifold where shape deformations are driven by the local concentration of proteins.
Additional mechanical properties, such as bending stiffness, in-plane elasticity, and volume or area constraints can be incorporated into our model by including further terms in the normal velocity.
While these additional features do not affect the instability criterion Eq.~\eqref{eq:secIII-lateral-instability-criterion-mapped}, they will introduce further nonlinearities, which in general will lead to complex pattern-forming dynamics and wavelength selection in the highly nonlinear regime.

\textit{Turing systems on growing domains.---}
Our model shows conceptual differences when compared to classical Turing models on homogeneously growing domains, such as uniformly growing planar lines~\cite{Maini.etal2012}.
In such systems, which grow uniformly in length, each length segment of the domain grows at the same rate, so that the dynamics of the metric can be eliminated by being absorbed into the temporal change of the total length.
For mass-conserving reaction-diffusion systems, this entails that the local total density is the only relevant degree of freedom in the system.
The system considered here involves a dynamic interplay between the local total density and the metric, which leads to (self-organized) non-uniform growth rates and thereby rich pattern-forming dynamics.
Classical Turing models have also been studied for non-uniformly growing lines~\cite{Krause.etal2021}, where, for example, one segment of the line is assumed to grow at a different rate from that of the remaining portion, which can effectively be described as a piecewise uniformly growing line.
It was found that this leads to asymmetric pattern formation and peak-splitting of patterns, which can be interpreted as regional patterns in analogy to our work. 
However, the underlying mechanism leading to such regional Turing patterns is, again, substantially different from our model.
In essence, Turing patterns in such systems occur (including peak-splitting) once the local line segment length exceeds a critical value, thus inducing (regional) Turing instabilities or frequency-doubling of the pattern.

Notably, these classical Turing models have been mainly studied in the quasi-stationary limit~\cite{Maini.etal2012,Krause.etal2021}, where one assumes that the pattern-forming dynamics unfolds on a much smaller time scale than domain growth.
While such an assumption is reasonable at larger scales, such as in the context of morphogenesis, the time scales of growth and pattern formation are generally not far apart in an intracellular context. 
This is evidenced by recent \textit{in vitro} experiments, which show that proteins are capable of dynamically deforming giant unilamellar vesicles (GUVs)~\cite{Litschel.etal2018}, or reshaping supported lipid bilayers~\cite{Rogez.etal2019}.
Therefore, here we have examined the full range of relative time scales (diffusive mass redistribution and shape deformations) by varying the Péclet number, and indeed found qualitative differences in the dynamics as a function of these time scales, such as a transition from traveling waves to standing waves. 
This underscores the relevance of the different time scales as an additional means by which mechanochemical patterns in cells may be controlled.
For concentration-dependent growth, as we have considered here, cells may achieve such control by regulating the total density of proteins.

\textit{Bulk-boundary coupling.---}
Protein patterns in biological systems often emerge at surfaces, such as the cell membrane, where proteins cooperatively bind to and detach from the membrane.
Consequently, proteins have to be transported from the bulk solution (cytosol) to the cell membrane, which is achieved by diffusive and advective fluxes in cells~\cite{Burkart.etal2022}.
This leads to cytosolic protein density gradients perpendicular to the membrane, and these gradients have been shown to be crucial for pattern formation in mass-conserving reaction--diffusion systems~\cite{Halatek.etal2018, Halatek.Frey2018,  Brauns.etal2021b, Wuerthner.etal2021}.
Another interesting extension of our work would be therefore to explicitly account for bulk-boundary coupling in the reaction--diffusion dynamics.
Potentially, this might yield additional interesting geometric effects, since shape deformations would (locally) alter the bulk-boundary ratio, which is an important control parameter for protein pattern formation~\cite{Thalmeier.etal2016, Halatek.Frey2018,Gessele.etal2020, Brauns.etal2021b,Feddersen.etal2021, Wuerthner.etal2021}.

\textit{Biologically realistic reaction networks.---}
We expect that our analysis can be transferred to more complex mass-conserving reaction--diffusion systems.
One prominent example is the Min protein system in \emph{E. coli}, which can generate a broad variety of self-organized patterns such as traveling waves, standing waves, chaos, and stationary patterns (for a review please refer to e.g. Ref.~\cite{Ramm.etal2019}).
Recently, it was shown that the \textit{in vitro} Min system in a heterogeneous setup (three-dimensional wedge-shaped geometry) leads to patterns on multiple length and time scales~\cite{Wuerthner.etal2021}. 
Importantly, these joint theoretical and experimental studies have shown that the large-scale dynamics can be characterized by diffusive redistribution of protein mass, which is the essential degree of freedom on large spatial and temporal scales.
In the present work, we found that spatial heterogeneities generally also occur in systems that exhibit a feedback loop between shape deformations and reaction--diffusion dynamics. 
Then, in contrast to systems with (fixed) spatially varying geometry as in the wedge setup mentioned above or in the context of a fixed cell shape, spatial heterogeneities and complex geometries are generated by the dynamics. 
One might therefore wonder why we do not observe multiscale patterns here.
The reason is that the two-component system has only one stable attractor (mesa or peak pattern)~\cite{Brauns.etal2020, Brauns.etal2021,Weyer.etal2022}, which significantly limits the phenomenology.
One could, however, readily apply our approach to the Min dynamics by replacing the reaction--diffusion component in our model with the biochemical reaction network of the Min system.
Coupling Min patterns to shape deformations may lead to interesting dynamics that possibly span multiple spatial and temporal scales, and the concept of regional instabilities would enable one to characterize and explain such multiscale patterns on dynamic manifolds.

Moreover, this could provide a rich field of research if one, for example, considers placing an additional lipid bilayer membrane at some height above a supported lipid bilayer membrane~\cite{Fu.etal2021}.
Now, if this additional lipid bilayer is not supported by a solid surface but is free standing, it can be deformed by the Min proteins and thereby dynamically affect the cytosolic volume between the two membranes and thus the local volume-boundary ratio. 
That Min proteins are indeed capable of deforming giant unilamellar vesicles was recently demonstrated experimentally~\cite{Litschel.etal2018,Fu.etal2021}.
We hypothesize that in such a system one could observe an intricate dynamic interplay between multiscale protein patterns and the dynamics of the free-standing membrane.

\begin{acknowledgments}
We would like to thank A. Ziepke and T. Roth for critical reading of the manuscript. This work was funded by the Deutsche Forschungsgemeinschaft (DFG, German Research Foundation) through the Collaborative Research Center (SFB) 1032 -- Project-ID 201269156 -- and the Excellence Cluster ORIGINS under Germany’s Excellence Strategy -- EXC-2094 -- 390783311. AG was supported by a DFG fellowship through the Graduate School of Quantitative Biosciences Munich (QBM). During his time at the Massachusetts Institute of Technology, AG was supported by the National Science Foundation (NSF) through grant number 2044895.
\end{acknowledgments}

\appendix

\section{Reaction term}
\label{appendix:reaction_term}

We adopted a reaction term that has been proposed as a conceptual model for cell polarization~\cite{Mori.etal2011,Brauns.etal2020}. 
The reaction kinetics are based on autocatalytic recruitment of membrane proteins and linear detachment:
\begin{equation}
    f(m,c) 
    = 
    \left[ k_\text{on} + k_\text{fb} \, \frac{m^2}{K_\text{d}^2 + m^2}\right] c - k_\text{off} \, m 
    \, .
    \label{eq:appendix-reaction-kinetics-static}
\end{equation}
For the specific parameters that we chose here (see Appendix~\ref{appendix:parameters}), we obtain an N-shaped nullcline, as qualitatively shown in Fig.~\ref{fig:phase-space}b.
Hence, for our choice of parameters, the reaction--diffusion model always produces mesa patterns (lower and upper plateau in the density profile which are connected by an interface, see Fig.~\ref{fig:phase-space}a), because the flux-balance subspace intersects the reactive nullcline at three points~\cite{Brauns.etal2020, Brauns.etal2021, Weyer.etal2022} (see Fig.~\ref{fig:phase-space}a,b).

The space- and time-dependent reaction term $\tilde{f}(x,t)$ for the rescaled densities $\widetilde{m}$ and $\tilde{c}$ then follows from Eq.~\eqref{eq:secIII-mapped-reaction} and takes the form:
\begin{equation}
  \tilde{f}(x,t)
  =
  \left[ k_\text{on} + k_\text{fb} \, \frac{\widetilde{m}^2}{K_\text{d}^2 \, g + \widetilde{m}^2}\right] \tilde{c} - k_\text{off} \, \widetilde{m} \, .
   \label{eq:appendix-mapped-reaction-kinetics}
\end{equation}

\section{Time-evolution of the curvature}
\label{appendix:time_evolution_curvature}

Instead of tracking the temporal change of the position vector, one may also study how the curvature at each point along the manifold evolves with time.
Since the curvature characterizes the (local) conformation of the one-dimensional manifold, one could also reconstruct the position vector from the solution of the curvature alone (up to translation and rotation)~\cite{Rogez.etal2019}.
Here, we present the derivation of the evolution equation for the curvature in the material frame $\mathcal{D}_t \kappa(\mat,t)$.

To this end, we first determine the commutator of time and arc length derivatives along the manifold:
\begin{align}
    \mathcal{D}_t \partial_s&=\mathcal{D}_t \left[\frac{1}{\sqrt{g(\mat,t)}} \partial_{\mat}\right]\nonumber\\
    &= \partial_s  \mathcal{D}_t + v_n \kappa \, \partial_s,
    \label{eq:appendix-derivatives-exchange}
\end{align}
where, after applying the chain rule, we used Eq.~\eqref{eq:secIII-comoving-timederiv-metric} to obtain the result above.
To proceed, we now use Eq.~\eqref{eq:appendix-derivatives-exchange} to determine the temporal evolution of the unit tangent vector:
\begin{align}
    \mathcal{D}_t \vec{\hat{\tau}}&=\mathcal{D}_t \partial_s \vec{r}=\partial_s \mathcal{D}_t \vec{r} + v_n \kappa \, \partial_s \vec{r}\nonumber\\
    &=\partial_s[v_n \vec{\hat{n}}] + v_n \kappa \, \vec{\hat{\tau}}\nonumber\\
    &=\partial_s v_n \vec{\hat{n}},
    \label{eq:appendix-time-evolution-tangent}
\end{align}
where we used the definition ${\mathcal{D}_t \vec{r}(\mat,t)=v_n \vec{\hat{n}}}$ and the fact that ${\partial_s \vec{\hat{n}}=-\kappa \vec{\hat{\tau}}}$.
Finally, by using Eqs.~\eqref{eq:appendix-derivatives-exchange} and~\eqref{eq:appendix-time-evolution-tangent} we compute the following expression:
\begin{align}
    \partial_s \mathcal{D}_t \vec{\hat{\tau}}&=\partial_s^2 v_n \vec{\hat{n}}+\partial_s v_n \partial_s \vec{\hat{n}}\nonumber\\
    &=\mathcal{D}_t \partial_s \vec{\hat{\tau}} - v_n \kappa \, \partial_s \vec{\hat{\tau}}\nonumber\\
    &=(\mathcal{D}_t \kappa-v_n \kappa^2)\vec{\hat{n}}+\kappa \, \mathcal{D}_t \vec{\hat{n}},
    \label{eq:appendix-time-evolution-curvature-derivation}
\end{align}
here, we used Eq.~\eqref{eq:curvature} to obtain the third line.
Comparing the first and last lines in the equation above, one finds that
\begin{equation}
    \mathcal{D}_t \kappa(\mat,t)= \partial_s^2 v_n + \kappa^2\,v_n.
    \label{eq:appendix-time-evolution-curvature}
\end{equation}

\section{Non-dimensionalized equations}
\label{appendix:nondimensionalized_equations}

After non-dimensionalization, we arrive at the following set of partial differential equations:
\begin{widetext}
\begin{subequations}
\begin{align}
    \partial_t \tilde{n}(x,t) 
    &= \partial_x \biggl[ \frac{\partial_x h}{g} \, v_y \, \tilde{n} + \frac{D}{\sqrt{g}} \, \partial_x\left(\frac{\tilde{n}}{\sqrt{g}}\right)
    + \frac{1 - D}{\sqrt{g}}\partial_x\left(\frac{\tilde{c}}{\sqrt{g}}\right)\biggr]  \, , \\
    \partial_t \tilde{c}(x, t) 
    &= \partial_x \left[ \frac{\partial_x h}{g} \, v_y \, \tilde{c} + \frac{1}{\sqrt{g}} \, \partial_x\left(\frac{\tilde{c}}{\sqrt{g}}\right) \right] 
    - \left[ k_\text{on} + k_\text{fb} \, \frac{(\tilde{n}-\tilde{c})^2}{K_\text{d}^2 \, g + (\tilde{n}-\tilde{c})^2}\right] \tilde{c} + k_\text{off} \, (\tilde{n}-\tilde{c}) \, , \\
    \partial_t h(x,t) &= v_y \, , \quad\text{where} \quad
    v_y = \tilde{\mu} \, \tilde{n} + \tilde{\gamma} \, \frac{1}{g} \, \partial_x^2 h \,.
\end{align}
\end{subequations}
\end{widetext}
Note that we solve here for the variables $\tilde{n}$ and $\tilde{c}$, instead of $\widetilde{m}$ and $\tilde{c}$ (which are related via local mass conservation $\tilde{n}=\widetilde{m}+\tilde{c}$, cf. Eq.~\eqref{eq:secIII-density-mapping}).
We solved these equations numerically with FENICs, which allowed us to perform the parameter sweeps with greater efficiency.
Furthermore, using two different softwares for solving the partial differential equations allowed us to further validate the accuracy and reliablity of our numerical results.

\section{Parameters}
\label{appendix:parameters}

For convenience, we have omitted physical units throughout the manuscript.
Here, we provide the values of the model parameters, and give an estimate of the typical length and time scales of protein patterns in biological systems.
\setlength{\tabcolsep}{15pt}
\begin{table}[htb]
\centering
\caption{Model parameters. If not otherwise specified, the parameter set below were used in this study.}
\begin{tabular}{@{} l  c  l @{}}
Parameter & Symbol & Value\\
  \hline
Cytosolic diffusion & $D_c$ & $\SI{0.1}{\micro m^2 \, s^{-1}}$ \\
Membrane diffusion & $D_m$ & $\SI{0.01}{\micro m^2 \, s^{-1}}$ \\
Average total density & $\langle n \rangle$ & $\SI{2.4}{\micro m^{-1}}$ \\ 
Attachment rate & $k_\text{on}$ & $\SI{0.07}{s^{-1}}$\\
Detachment rate & $k_\text{off}$ & $\SI{1.0}{s^{-1}}$\\
Recruitment rate  & $k_\mathrm{fb}$ & $\SI{1.0}{s^{-1}}$\\
Carrying capacity  & $K_d$ & $\SI{1.0}{\micro m^{-1}}$\\
Coupling strength  & $\mu$ & $\SI{0.05}{\micro m^2 \, s^{-1}}$\\
Line tension & $\gamma$ & $\SI{0.001}{\micro m^2 \, s^{-1}}$\\
\hline
\end{tabular}
\label{tab:parameters}
\end{table}
The typical system size in an intracellular context is $L_0 \approx \SI{10}{\micro m}$. The typical value for membrane diffusion is $D_m \sim \SI{0.01}{\micro m^2 \, s^{-1}}$, while in the cytosol $D_c \sim \SIrange[range-phrase=-,range-units=single]{0.1}{10}{\micro m^2 \, s^{-1}}$.
The characteristic time scale of pattern formation is determined by the kinetic parameters as well as mass redistribution in the cytosol and on the membrane (via diffusion and possibly advection), and is typically on the order of minutes in an intracellular context~\cite{Burkart.etal2022}.
In this work, length scales are given in units of $\SI{1}{\micro m}$, and time scales in units of $k_\text{off}= \SI{1.0}{s^{-1}}$ (see Table~\ref{tab:parameters}).

\section{Linear stability analysis for a one-component system in the absence of chemical reactions}
\label{appendix:lsa_1}

To gain further insight into how geometry deformations affect the relaxation of a single membrane-bound particle species to a homogeneous state via diffusion, we consider the following simplified model:
\begin{subequations}
\begin{align}
    \mathcal{D}_t \varrho(\mat,t) 
    &= \frac{1}{\sqrt{g}} \frac{\partial}{\partial\mat} \biggl[ \frac{D}{\sqrt{g}} \frac{\partial \varrho}{\partial\mat} \biggr] + \kappa \, v_n \, \varrho \, , \\
    \mathcal{D}_t \kappa(\mat,t) 
    &= \kappa^2 \, v_n + \frac{1}{\sqrt{g}} \frac{\partial}{\partial\mat} \biggl[ \frac{1}{\sqrt{g}} \frac{\partial v_n}{\partial\mat} \biggr] \, , \text{and} \\
    \mathcal{D}_t g(\mat,t) 
    &= - 2 \, g \, \kappa \, v_n \,, \quad\text{where}\quad v_n = \mu \varrho \,.
\end{align}
\label{eq:supp:reduced_model}%
\end{subequations}
We perform a linear stability analysis around a homogeneous steady state, $\varrho = \varrho^* + \delta \varrho$, with a flat configuration of the interface, $\kappa = \delta \kappa$ and $g = g^* + \delta g$.
Then, up to linear order, Eqs.~\eqref{eq:supp:reduced_model} further simplify to:
\begin{subequations}
\begin{align}
    \mathcal{D}_t [\delta \varrho(\mat,t)] 
    &= \frac{D}{g^*} \partial_{\mat}^2 [\delta \varrho] + \mu \, {\varrho^*}^2 \, [\delta\kappa] \, , \\
    \mathcal{D}_t [\delta\kappa(\mat,t)] 
    &= \frac{\mu}{g^*} \partial_{\mat}^2 [ \delta \varrho ] \, .
\end{align}
\label{eq:supp:reduced_model_lsa}%
\end{subequations}
Note that we have here omitted the dynamics of the metric $g$, since it decouples from the set of equations~\eqref{eq:supp:reduced_model_lsa} to linear order and is therefore not relevant.
Taking the Fourier transform of the perturbations,
\begin{subequations}
\begin{align}
    \delta \varrho(\mat,t) 
    &= \frac{1}{2\pi} \int dq \, \delta \hat{\varrho}(q, t) \, \exp(i \, q \, \mat) 
    \,, \\
    \delta \kappa(\mat,t) 
    &= \frac{1}{2\pi} \int dq \, \delta \hat{\kappa}(q, t) \, \exp(i \, q \, \mat) 
\end{align}
\end{subequations}
we thus arrive at:
\begin{align}
    \mathcal{D}_t 
    \begin{bmatrix}
    \delta \hat{\varrho}(q,t) \\
    \delta \hat{\kappa}(q,t)
    \end{bmatrix} 
    &=
    \begin{bmatrix}
    - D q^2 / g^* & \mu {\varrho^*}^2 \\
    - \mu q^2 / g^* & 0
    \end{bmatrix} 
    \cdot
    \begin{bmatrix}
    \delta \hat{\varrho}(q,t) \\
    \delta \hat{\kappa}(q,t)
    \end{bmatrix} 
    \nonumber \\
    &\coloneqq
    \vec{J}
    \cdot
    \begin{bmatrix}
    \delta \hat{\varrho}(q,t) \\
    \delta \hat{\kappa}(q,t)
    \end{bmatrix} 
    \, ,
\end{align}
where we have lastly defined the Jacobian $\vec{J}$ of the linearized system.
Note that the trace of the Jacobian is always negative, $\tr\vec{J} = -D q^2/g^* <0$ while its determinant is always positive, $\det\vec{J} = \mu^2 q^2 {\varrho^*}^2 / g^* > 0$.
Thus, the system is always stable.
\begin{figure}[tbh]
    \centering
    \includegraphics[width=0.5\textwidth]{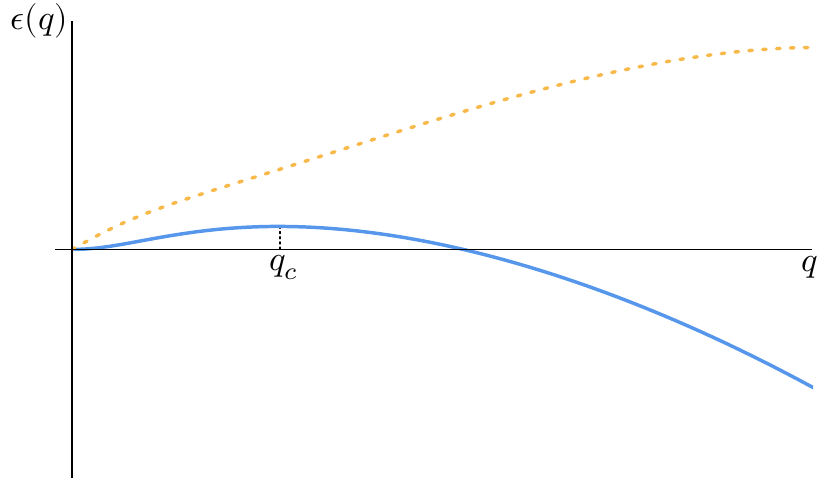}
    \caption{Typical dispersion relation for the two-component model on a dynamic one-dimensional manifold.
    The blue solid line shows the real part of the growth rate $\epsilon(q)$, and the orange dashed line shows the imaginary part.
    The fact that the imaginary part is non-zero indicates local oscillations that lead to traveling wave patterns.}
    \label{fig:appendix_dispersion_relation}
\end{figure}
We find that all slow modes below a cricitcal wave number,
\begin{equation}
    \frac{q^2}{g^*} < \frac{q_c^2}{g^*} \coloneqq \left(2\frac{\mu \, \varrho^*}{D}\right)^2 \, ,
\end{equation}
are stable spirals, while all fast modes are stable nodes.

\section{Linear stability analysis for the two-component system with mechanochemical coupling}
\label{appendix:lsa_2comp}
We can now extend the analysis in Appendix~\ref{appendix:lsa_1} to the two-component system, where the Jacobian in this case is given by:
\begin{widetext}
\begin{equation}
    \vec{J} = 
    \begin{bmatrix}
    - D_m q^2 / g^* + \partial_m f & \partial_c f & \mu \, m^* (m^*+c^*) \\
    - \partial_m f & - D_c q^2 / g^* - \partial_c f & \mu \, c^* (m^*+c^*) \\
    - \mu q^2 / g^* & - \mu q^2 / g^* & 0
    \end{bmatrix}, 
    \label{eq:appendix-jacobian-two-comp}
\end{equation}
\end{widetext}
with $\partial_{m/c} f:=\left.\partial_{m/c} f\right|_{[m^*,c^*]}$.
From~\eqref{eq:appendix-jacobian-two-comp} we determined the dispersion relation $\epsilon(q)$ which relates the growth rate of perturbations to the mode number $q$ (Fig.~\ref{fig:appendix_dispersion_relation}).
While the growth rate of the two-component model on a static planar geometry contains only a real part in the unstable regime~\cite{Brauns.etal2020}, we find here that both the real and imaginary part of the growth rate can become positive.
Hence, this suggests that the system exhibits traveling wave patterns, since a non-zero imaginary part indicates local oscillations, as confirmed by our simulations.
From~\eqref{eq:appendix-jacobian-two-comp} we further numerically determined the fastest growing mode $q_c$ (which corresponds to the eigenvalue with the largest real part, see Fig.~\ref{fig:appendix_dispersion_relation}), from which we obtained an estimate for the initial pattern wavelength in our simulations $\lambda_c \approx \SI{4}{\micro m}$ (using the parameters provided in Table~\ref{tab:parameters}).

\bibliography{library}

\end{document}